\documentclass[aps,prd,preprintnumbers,twocolumn,superscriptaddress,nofootinbib]{revtex4-1}
\usepackage{lmodern}
\usepackage{hyperref}
\hypersetup{colorlinks=true,linkcolor=purple,anchorcolor=blue,citecolor=blue, filecolor=blue,urlcolor=blue,bookmarksnumbered=true,
pdfview=FitB
}
\usepackage{mathrsfs}
\usepackage{rsfso}
\usepackage{color}
\usepackage{xcolor}
\usepackage{ulem}
\usepackage{csquotes}
\colorlet{purple1}{blue!70!red}
\colorlet{darkred}{red!50!black}

\usepackage{graphicx}
\usepackage[bf,SL,BF]{subfigure}
\usepackage{psfrag}
\usepackage{color}
\usepackage{amssymb}
\usepackage{amsmath}
\usepackage{epstopdf}
\usepackage{natbib}
\usepackage{amssymb}
\usepackage{amsmath,amssymb}
\usepackage{mathtools}
\usepackage{float}
\usepackage{color}
\usepackage{leftidx}
\usepackage{booktabs}
\usepackage{latexsym}
\usepackage{revsymb}
\usepackage{multirow}
\usepackage{hypcap}
\usepackage{array}
\usepackage[english]{babel}
\usepackage{amsmath}
\usepackage{cleveref}
\def\orcid#1{\kern .08em\href{https://orcid.org/#1}{\includegraphics[keepaspectratio,width=0.7em]{ORCID_iD.png}}}
\newcommand{\be}{\begin{eqnarray}}
	\newcommand{\ee}{\end{eqnarray}}
	
\def\orcid#1{\kern .08em\href{https://orcid.org/#1}{\includegraphics[keepaspectratio,width=0.7em]{ORCID_iD.png}}}
\newcommand{\kp}{k_{\perp}}
\newcommand{\qp}{q_{\perp}}
\newcommand{\bfq}{{\bf q}_{\perp}}

\newcommand{\bfp}{{\bf p}_{\perp}}
\newcommand{\bfk}{{\bf k}_{\perp}}

\newcommand{\bfP}{{\bf P}_{\perp}}

\newcommand{\ba}{\begin{align}}  
\newcommand{\ea}{\end{align}}  

\usepackage{braket}

\usepackage{comment}
\begin{document}

	\title{Gluon gravitational form factors of the proton in a light-front spectator model}	
	
	\author{Amrita~Sain}
    \email{amrita@impcas.ac.cn}
	\affiliation{Institute of Modern Physics, Chinese Academy of Sciences, Lanzhou 730000, China}
	\affiliation{School of Nuclear Science and Technology, University of Chinese Academy of Sciences, Beijing 100049, China}

        \author{Poonam~Choudhary}
    \email{poonamch@iitb.ac.in} 
	\affiliation{Department of Physics, Indian Institute of Technology Bombay, Powai, Mumbai 400076, India}

	\author{Bheemsehan~Gurjar}
	\email{gbheem@iitk.ac.in} 
	\affiliation{Department of Physics, Indian Institute of Technology Kanpur, Kanpur-208016, India}
	
	\author{Chandan~Mondal}
		\email{mondal@impcas.ac.cn} 
	\affiliation{Institute of Modern Physics, Chinese Academy of Sciences, Lanzhou 730000, China}
	\affiliation{School of Nuclear Science and Technology, University of Chinese Academy of Sciences, Beijing 100049, China}

     	\author{Dipankar~Chakrabarti}
	\email{dipankar@iitk.ac.in} 
	\affiliation{Department of Physics, Indian Institute of Technology Kanpur, Kanpur-208016, India}

    \author{Asmita~Mukherjee}
	\email{asmita@phy.iitb.ac.in} 
	\affiliation{Department of Physics, Indian Institute of Technology Bombay, Powai, Mumbai 400076, India}

	\date{\today}

\begin{abstract}
We calculate the gluon gravitational form factors (GFFs) of the proton using a light-front spectator model based on soft-wall AdS/QCD, where the active parton is a gluon. The model parameters are determined by fitting the unpolarized gluon distribution function to the NNPDF3.0nlo dataset. Subsequently, we predict the polarized gluon distribution, finding consistency with global analyses. Our predictions for the gluon GFFs show good agreement with recent lattice QCD simulations and experimental extractions. Using these gluon GFFs, we compute the contribution of the gluon to the proton mass and mechanical radii. We also analyze the gluonic contribution to the two-dimensional Galilean densities, including energy density, radial and tangential pressure, isotropic pressure, and pressure anisotropy of the proton.
\end{abstract}

\maketitle

\section{Introduction}\label{intro}
Understanding the structure of the proton, a bound state of quarks and gluons governed by quantum chromodynamics (QCD), remains a major challenge in nuclear and particle physics~\cite{Accardi:2012qut,AbdulKhalek:2021gbh, Anderle:2021wcy,Burkert:2018nvj,Accardi:2023chb}. The mechanical properties of the proton, such as the distribution of mass, spin, and pressure among quarks and gluons, are of great interest~\cite{Lorce:2018egm, Polyakov:2002yz, Polyakov:2018zvc, Burkert:2023wzr}. These properties are linked to the gravitational form factors (GFFs), defined by the matrix elements of the QCD energy-momentum tensor (EMT) in the proton state, highlighting their fundamental importance.

The GFFs associated with the symmetric, traceless part of the EMT correspond to the second Mellin moments of the generalized parton distributions (GPDs)~\cite{Diehl:2003ny, Ji:1996nm, Radyushkin:1997ki}, allowing them to be constrained by experimental data from deeply virtual Compton scattering (DVCS)~\cite{Ji:1996ek, dHose:2016mda, Kumericki:2016ehc} and meson production~\cite{Radyushkin:1996ru, Collins:1996fb}. Proton's quark GFFs have been extracted from DVCS measurements at Jefferson Lab (JLab)~\cite{Burkert:2018bqq}, and further constraints can be refined using GPDs accessible at current and future facilities, including the PANDA experiment at Facility for Antiproton and Ion Research (FAIR)~\cite{PANDA:2009yku}, proposed Electron-Ion Colliders (EICs) \cite{AbdulKhalek:2021gbh, Anderle:2021wcy}, Nuclotron-based Ion Collider Facility (NICA)~\cite{MPD:2022qhn}, International Linear Collider (ILC), and Japan proton Accelerator Complex (J-PARC)~\cite{Kumano:2022cje}. Significant progress has been made in the theoretical determination of the proton's GFFs through lattice QCD, phenomenology, and QCD-inspired models~\cite{Chen:2001pva,Belitsky:2002jp,Dorati:2007bk,Schweitzer:2002nm,Wakamatsu:2006dy,Goeke:2007fq,Goeke:2007fp,Wakamatsu:2007uc,Neubelt:2019sou,Cebulla:2007ei,Kim:2012ts,Anikin:2019kwi,Chakrabarti:2015lba,Chakrabarti:2020kdc,Chakrabarti:2021mfd,Chakrabarti:2016mwn,Kumar:2017dbf,Pasquini:2014vua,Polyakov:2018exb,Abidin:2009hr,Mondal:2015fok,Choudhary:2022den,Hagler:2003jd,Gockeler:2003jfa,LHPC:2010jcs,LHPC:2007blg,QCDSF-UKQCD:2007gdl,Deka:2013zha,Cao:2023ohj,Hatta:2018sqd,Tanaka:2018nae,Nair:2024fit,Wang:2023fmx,Mondal:2016xsm,Cao:2024zlf}. 

However, the gluon contributions to the proton's GFFs remain poorly constrained. These experiments can hardly observe the gluon. Near-threshold charmonium and bottomonium photo or leptoproduction offers a unique probe of the proton's gluonic structure~\cite{Mamo:2019mka,Mamo:2022eui,Guo:2023qgu,Hatta:2018ina,Boussarie:2020vmu,Pentchev:2024sho,GlueX:2019mkq,GlueX:2023pev}. Future experiments at EICs aim to explore these gluonic contributions~\cite{AbdulKhalek:2021gbh,Anderle:2021wcy}, while JLab has already conducted near-threshold $J/\psi$ production studies, revealing promising results~\cite{GlueX:2019mkq,Duran:2022xag,GlueX:2023pev,Liu:2024yqa,Hechenberger:2024abg}. This process is expected to be dominated by two-gluon exchange due to the charm quark's high mass and proximity to threshold~\cite{Brodsky:1997gh,Sun:2021gmi}. Under factorization, with the charmonium mass as the hard scale, the process involves the proton's gluonic GPDs, which can be related to the gluon GFFs through graviton-like exchange~\cite{Guo:2021ibg,Hatta:2021can}. Experimental validation of these assumptions is essential to fully understand the gluonic structure of the proton. In Ref.~\cite{Duran:2022xag}, the proton's gluon GFFs were extracted by fitting the $J/\psi$-007 data using theoretical predictions from holographic~\cite{Mamo:2021krl} and GPD~\cite{Guo:2021ibg} approaches, focusing on low $|t|$ cross-section data, where $|t|$ defines square of the total momentum transfer. The GPD analysis in Ref.~\cite{Guo:2023pqw} combined data from GlueX and $J/\psi$-007, extracting the gluon GFFs through an expansion in the skewness parameter $\xi$ as $\xi \to 1$, which is valid for high $|t|$ due to the kinematic link between $t$ and $\xi$~\cite{Hatta:2021can}. Both studies model the gluon GFFs using dipole/tripole functions, analogous to electromagnetic form factors as proposed in Ref.~\cite{Frankfurt:2002ka}, and parameterized using lattice QCD results~\cite{Pefkou:2021fni,Hackett:2023rif}.

Theoretical investigations into gluon GFFs have been less extensive compared to those of quark GFFs. Most phenomenological models for the nucleon do not include gluonic degrees of freedom, limiting the focus to quark GFFs. However, certain GFFs, such as the { $D$}-term, which contributes to pressure and shear force distributions in the nucleon, depend on the ``bad" components of the energy-momentum tensor~\cite{Hu:2024edc} that include quark-gluon interactions. This underscores the importance of studying the role of gluons in these distributions. Despite this, theoretical studies and experimental constraints on gluon GFFs remain limited.
The gluon GFFs have so far only been studied in  lattice QCD calculations~\cite{Shanahan:2018nnv,Hackett:2023rif,Shanahan:2018pib,Pefkou:2021fni,Alexandrou:2020sml,Yang:2018bft,Yang:2018nqn,Alexandrou:2017oeh}, holographic QCD~\cite{Mamo:2021krl,Mamo:2019mka,Mamo:2022eui}, an extended holographic light-front QCD
framework~\cite{deTeramond:2021lxc,Gurjar:2022jkx}, dressed quark model~\cite{More:2023pcy} and more recently using Bethe-Salpeter equation approach~\cite{Yao:2024ixu}. Improved QCD constraints on the proton's gluon GFFs are crucial, as they help refine target kinematics for experiments and provide theoretical predictions to compare with future experimental results. 

The GFFs encode spatial densities—such as momentum density, energy density, pressure, and shear density, among others~\cite{Xu:2024cfa}—through Fourier transforms. However, it is crucial to carefully interpret the physical meaning of these densities. It has been established that fully relativistic densities can only be meaningfully obtained via two-dimensional (2D) Fourier transforms at fixed light-front time~\cite{Burkardt:2002hr,Miller:2007uy,Miller:2009sg,Carlson:2007xd,Miller:2018ybm,Freese:2021czn,Xu:2024cfa,Mondal:2017lph}. In contrast, the three-dimensional (3D) Breit frame density relies on the incorrect assumption that the hadron can be spatially localized~\cite{Miller:2018ybm,Jaffe:2020ebz}. A detailed comparison of density definitions across different frames is provided in Refs.~\cite{Lorce:2018egm,Freese:2022fat}. Due to the Galilean symmetry inherent in the light-front framework, the resulting 2D distributions remain fully relativistic. The relationship between 2D { light-front} and 3D { Breit frame} distributions can be established via the Abel transformation, as discussed in Ref.~\cite{Panteleeva:2021iip}. In this context, the gluonic contributions to mechanical properties such as pressure, shear, and energy distributions of a dressed quark state have been investigated in Ref.~\cite{More:2023pcy}. { Gluon GFFs and associated distributions are fundamental for understanding the mechanical stability of the hadron. The pressure and shear force distributions inside a nucleon coming from the gluon content of the proton in the Breit frame have been investigated on the lattice~\cite{Shanahan:2018nnv} and Bethe–Salpeter approach~\cite{Yao:2024ixu}}.{ Additionally, Ref. \cite{Pefkou:2021fni} discusses { gluonic contribution} to various densities, including pressure and shear distributions, in both the Breit frame and the infinite momentum frame.} 

In this work, we employ a recently developed spectator model~\cite{Chakrabarti:2023djs, Chakrabarti:2024hwx} that incorporates an active gluon to study the proton’s gluon GFFs. The proton is modeled as a composite system with an active gluon and a spin-$\frac{1}{2}$ spectator, primarily composed of three valence quarks at low energies. The light-front wave functions (LFWFs) are constructed from two-particle effective wave functions of soft-wall AdS/QCD~\cite{Brodsky:2014yha}. The model parameters are determined by fitting the unpolarized gluon PDF to the NNPDF3.0nlo global analysis. Using these LFWFs, the gluon helicity PDF and GFFs are subsequently derived as model predictions.  
We calculate the gluon GFFs \( A(Q^2) \), \( B(Q^2) \), \( \overline{C}(Q^2) \) and \( D(Q^2) \) from the gluon part of the EMT, finding good consistency with recent lattice QCD simulations~\cite{Hackett:2023rif,Pefkou:2021fni} and experimental extractions~\cite{Duran:2022xag,Guo:2021ibg}. Using these GFFs, we analyze the {{2D}} light-front gluon {{densities}}  within the proton, including the pressure and shear forces, as well as its mass and mechanical radii. { Further, We also compute the 2D light-front Galilean energy density, isotropic pressure and pressure anisotropy.} 
\section{Light-front gluon-spectator model revisited}
We adopt the generic ansatz of the light-front gluon-spectator model for the proton~\cite{Lu:2016vqu,Chakrabarti:2023djs,Lyubovitskij:2020xqj}, where the light-front wave functions are derived from the effective solution of soft-wall AdS/QCD. In this model, the proton is treated as a system consisting of an active gluon and a composite spectator state, with the spectator assumed to have spin $\frac{1}{2}$~\cite{Lu:2016vqu}. The two-particle Fock-state expansion for the proton's spin components, $J^z = \pm \frac{1}{2}$, in a frame where the proton's transverse momentum vanishes, i.e., $P \equiv \big(P^+, \textbf{0}_\perp, \frac{M^2}{P^+}\big)$, is expressed as
	\begin{align}\label{state}\nonumber
		&|P;\uparrow(\downarrow)\rangle
		= \int \frac{\mathrm{d}^2 {\bf k}_\perp \mathrm{d} x}{16 \pi^3 \sqrt{x(1-x)}}\nonumber\\ 
		&\times \Bigg[\psi_{+1+\frac{1}{2}}^{\uparrow(\downarrow)}\left(x, {\bf k}_\perp\right)\left|+1,+\frac{1}{2} ; x P^{+}, {\bf k}_\perp\right\rangle\nonumber\\ 
		&
		+\psi_{+1-\frac{1}{2}}^{\uparrow(\downarrow)}\left(x, {\bf k}_\perp \right)\left|+1,-\frac{1}{2} ; x P^{+}, {\bf k}_\perp \right\rangle\nonumber\\ 
		&+\psi_{-1+\frac{1}{2}}^{\uparrow(\downarrow)}\left(x, {\bf k}_\perp \right)\left|-1,+\frac{1}{2} ; x P^{+}, {\bf k}_\perp \right\rangle\nonumber\\ 
		&+\psi_{-1-\frac{1}{2}}^{\uparrow(\downarrow)}\left(x, {\bf k}_\perp\right)\left|-1,-\frac{1}{2} ; x P^{+}, {\bf k}_\perp\right\rangle\Bigg].
	\end{align}	
For nonzero transverse momentum of the proton, i.e., $\bfP \neq 0$, the physical transverse momenta of the gluon and spectator are given by $\bfp^g = x\bfP + \bfk$ and $\bfp^s = (1-x)\bfP - \bfk$, respectively, where $\bfk$ represents the relative transverse momentum of the gluon. The LFWFs, denoted by $\psi_{\lambda_g,\lambda_s}^{\lambda_p}(x,\bfk)$ with the proton's helicity $\lambda_p = \uparrow (\downarrow)$, correspond to the two-particle state $|\lambda_{g}, \lambda_s; xP^{+}, \bfk \rangle$, with the gluon's helicity $\lambda_g = \pm 1$, and the spectator's helicity $\lambda_s = \pm \frac{1}{2}$. The LFWFs in Eq.~(\ref{state}) are inspired by the physical electron's wave functions~\cite{Brodsky:2000ii}, which consist of a spin-1 photon and a spin-$\frac{1}{2}$ electron.

The LFWFs of the proton with $J_{z}=+\frac{1}{2}$ at the scale $\mu_0=2$ GeV are given by~\cite{Chakrabarti:2023djs},
	\begin{eqnarray} \label{LFWFsuparrow}   \nonumber
		\psi_{+1+\frac{1}{2}}^{\uparrow}\left(x,{\bf k}_\perp\right)&=&-\sqrt{2}\frac{(-k^{(1)}_{\perp}+ik^{(2)}_{\perp})}{x(1-x)}\varphi(x,{\bf k}_\perp^2), \\ \nonumber
		\psi_{+1-\frac{1}{2}}^{\uparrow}\left(x, {\bf k}_\perp\right)&=&-\sqrt{2}\bigg( M-\frac{M_s}{(1-x)} \bigg) \varphi(x,{\bf k}_\perp^2), \\ \nonumber
		\psi_{-1+\frac{1}{2}}^{\uparrow}\left(x, {\bf k}_\perp\right)&=&-\sqrt{2}\frac{(k^{(1)}_{\perp}+ik^{(2)}_{\perp})}{x}\varphi(x,{\bf k}_\perp^2), \\
		\psi_{-1-\frac{1}{2}}^{\uparrow}\left(x, {\bf k}_\perp\right)&=&0,
	\end{eqnarray}
while for the proton with $J_{z}=-\frac{1}{2}$, the LFWFs are expressed as: 
	\begin{eqnarray} \label{LFWFsdownarrow}   \nonumber
		\psi_{+1+\frac{1}{2}}^{\downarrow}\left(x, {\bf k}_\perp\right)&=& 0, \\ \nonumber
		\psi_{+1-\frac{1}{2}}^{\downarrow}\left(x,{\bf k}_\perp\right)&=&-\sqrt{2}\frac{(-k^{(1)}_{\perp}+ik^{(2)}_{\perp})}{x}\varphi(x,{\bf k}_\perp^2), \\ \nonumber
		\psi_{-1+\frac{1}{2}}^{\downarrow}\left(x, {\bf k}_\perp\right)&=&-\sqrt{2}\bigg( M-\frac{M_s}{(1-x)} \bigg) \varphi(x,{\bf k}_\perp^2),  \\
		\psi_{-1-\frac{1}{2}}^{\downarrow}\left(x, {\bf k}_\perp \right)&=& -\sqrt{2}\frac{(k^{(1)}_{\perp}+ik^{(2)}_{\perp})}{x(1-x)}\varphi(x,{\bf k}_\perp^2).
	\end{eqnarray}
Here, $M$ and $M_s$ denote the masses of the proton and spectator, respectively. The function $\varphi(x,\bfk^2)$ is a modified version of the soft-wall AdS/QCD wave function~\cite{Brodsky:2014yha,Gutsche:2013zia,Chakrabarti:2023djs}, modeled with the introduction of parameters $a$ and $b$, 
\begin{align}\label{AdSphi}
\varphi(x,{\bf k}_\perp^2)=&N_{g}\frac{4\pi}{\kappa}\sqrt{\frac{\log[1/(1-x)]}{x}}x^{b}(1-x)^{a}\nonumber\\
&\times\exp{\Big[-\frac{\log[1/(1-x)]}{2\kappa^{2}x^2}{\bf k}_\perp^{2}\Big]},
\end{align}
with an emerging scale, $\kappa$, which governs the transverse dynamics of the gluon. The parameters $a$ and $b$ control the asymptotic behavior of the gluon PDFs~\cite{Brodsky:1989db,Brodsky:1994kg}, and, along with the normalization constant $N_g$, are fit to the gluon unpolarized PDF using NNPDF3.0 data at the scale $\mu_0 = 2$ GeV~\cite{Chakrabarti:2023djs}. The scale parameter employed in Refs.~\cite{Chakrabarti:2023djs,Chakrabarti:2024hwx}: $\kappa = 0.4$ GeV, was determined from quark dynamics to fit the proton electromagnetic form factors~\cite{Chakrabarti:2013gra}. This led to discrepancies in the gluon GFFs $A_{g}(Q^2)$ and $B_{g}(Q^2)$ compared to lattice QCD and phenomenological fits~\cite{Chakrabarti:2024hwx}. It suggests that gluon dynamics differs from quarks. In this work, we use a different value of $\kappa = 2.62$ GeV for better alignment with gluon GFFs while maintaining consistent gluon PDFs. To ensure proton stability, the spectator mass is set to $M_s = 0.985_{-0.045}^{+0.044}$ GeV~\cite{Chakrabarti:2023djs}, larger than the proton mass and the gluon mass is taken as $M_g = 0$. 
The model parameters are summarized in Table~\ref{Tab:modelparameters}.

\begin{table}[ht]
\caption{Parameters of the model.}
\centering
\begin{tabular}[t]{lcccc}
\toprule\hline
~~$\kappa$~~ & ~~$a$~~ & ~~$b$~~\\
\hline
~~$2.62~\mathrm{GeV}$~~ &~~ $3.880 \pm 0.223$ ~~&~~ $-0.530 \pm 0.007$~~  \\
\hline\hline
\end{tabular}
\label{Tab:modelparameters}
\end{table}
\section{Gluon PDFs}
The LFWF overlap representation of gluon unpolarized and helicity PDFs are given as,
	\begin{align} \label{unpolTMDoverlap}
		f_{1}^{g}(x)=&\int \frac{{\rm d}^2\bfk}{16\pi^{3}}
		\Big[|\psi_{+1 +\frac{1}{2}}^{\uparrow}(x,\bfk)|^{2}\nonumber\\
		&+|\psi_{+1 -\frac{1}{2}}^{\uparrow}(x,\bfk)|^{2}+|\psi_{-1 +\frac{1}{2}}^{\uparrow}(x,\bfk)|^{2} \Big],\\
				\Delta g(x)=
		&\int \frac{{\rm d}^2\bfk}{16\pi^{3}}
		\Big[|\psi_{+1 +\frac{1}{2}}^{\uparrow}(x,\bfk)|^{2}\nonumber\\
		&+|\psi_{+1 -\frac{1}{2}}^{\uparrow}(x,\bfk)|^{2}-|\psi_{-1 +\frac{1}{2}}^{\uparrow}(x,\bfk)|^{2} \Big].
	\end{align}

\begin{figure}[htp]
\centering
\includegraphics[width=0.44\textwidth]{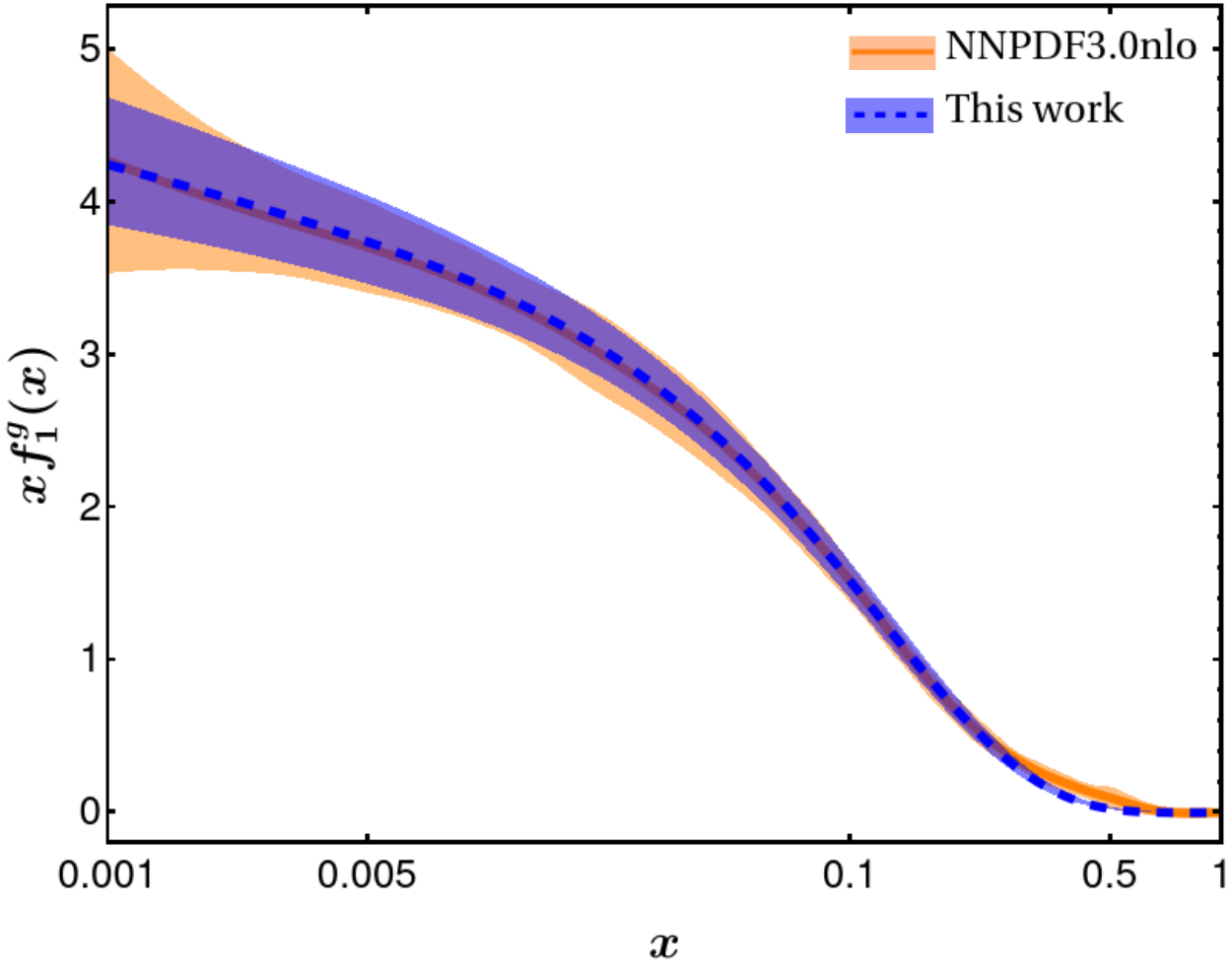}
\caption{\label{f1g} Our unpolarized gluon PDF $f_1^g(x)$ (blue dashed line with blue band corresponding to the uncertainty in model parameters). compared with the NNPDF3.0nlo data set (orange band)~\cite{Ball:2017otu} in the kinematics region $0.001 \leq x \leq 1 $ at $\mu_0=2$~GeV.}
\end{figure}

Fig.~\ref{f1g} presents our fit for the unpolarized gluon distribution $xf_{1}^{g}(x)$ at $\mu_{0} = 2$ GeV with a normalization constant $N_g = 0.32$. We estimate uncertainties on the parameters $a$ and $b$ that account for the model selections and major fitting uncertainties. A $2\sigma$ uncertainty is assumed as the standard maximum error for further analysis. The solid orange band represents the NNPDF3.0 parametrization of $xf_{1}^{g}(x)$~\cite{Ball:2017otu}, while the blue dashed line with blue band indicate our model results within the $2\sigma$ error range. Using those fitted model parameters as summarized in Table~\ref{Tab:modelparameters}, we then present the gluon helicity PDF and the GFFs. We also compute the mechanical properties, such as pressure and shear force distributions, as well as mechanical and mass radii.

The average longitudinal momentum carried by the gluon is defined as the second Mellin moment of the unpolarized PDF, expressed as,
$\langle x\rangle_{g}=\int_{0}^{1}{\rm d}x\, x f_{1}^{g}(x).$
We obtain $\langle x\rangle_{g} = 0.404^{+0.037}_{-0.039}$, which is in close agreement with the value obtained from the
global fit: $\langle x\rangle_{g} = 0.414\pm0.008$~\cite{Hou:2019efy} and different lattice QCD calculations at $\mu = 2$ GeV~\cite{Alexandrou:2020sml,Pefkou:2021fni,Yang:2018nqn,Hackett:2023rif}.

\begin{figure}[htp]
\centering
\includegraphics[width=0.44\textwidth]{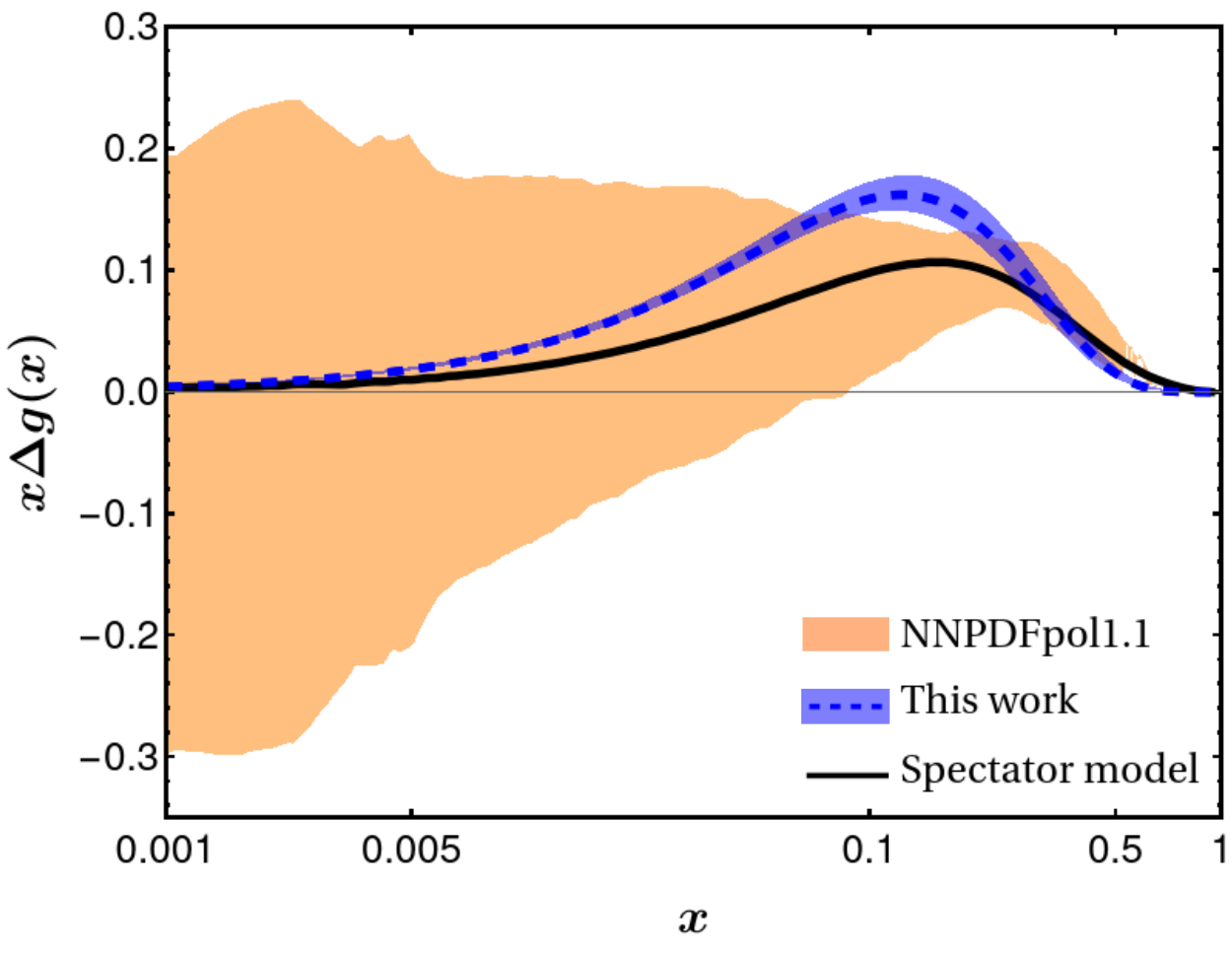}
\caption{\label{g1lg} Our gluon helicity PDF, $\Delta g(x)$ (blue dashed line with blue band corresponding to the uncertainty in model parameters) compared with the NNPDFpol1.1 (orange band)~\cite{Nocera:2014gqa}  and the spectator model results (black solid line)~\cite{Bacchetta:2020vty} at $\mu_{0} = 2$ GeV.}
\end{figure}

In Fig.~\ref{g1lg}, we present the intrinsic nonperturbative gluon helicity distribution, $x \Delta g(x)$, at the initial model scale $\mu_0 = 2$~GeV, comparing our results with the global analysis from the NNPDFpol1.1 Collaboration~\cite{Nocera:2014gqa} and another spectator model by Bacchetta {\it et al.}~\cite{Bacchetta:2020vty}. Our prediction for the proton's gluon helicity PDF shows good agreement with the global fit. The uncertainty band arises from variations in the model parameters, $a$, $b$ and $M_{s}$. Notably, at the model scale, the percentage uncertainty in $x \Delta g(x)$ is smaller than that in $x f_1(x)$. However, global analyses exhibit significant uncertainties, making $\Delta g(x)$ poorly constrained, especially regarding the sign, both in the small and large $x$ regions.

The gluon spin contribution, $\Delta G$, to the proton spin is determined by the first moment of the gluon helicity PDF, $\Delta g(x)$. Our analysis predicts $\Delta G = 0.42^{+0.04}_{-0.04}$ for $x_g \in [0.001, 1]$, which is significant for the proton spin. Table~\ref{Tab:spincontributions} shows the dependence of gluon helicity on the $x$ range, compared with experimental data and global fits. We find that the maximum contribution to gluon helicity arises from the small-$x$ region, with our model predicting relatively larger contributions across $x$ ranges. Specifically, our model predicts a larger gluon spin contribution, $\Delta G = 0.33\pm 0.03$, compared to recent analyses using updated datasets and PHENIX measurements~\cite{PHENIX:2008swq}, which report $\Delta G = 0.2$ for $x_g \in [0.02, 0.3]$. For $x_g \in [0.05, 0.2]$, our prediction, $\Delta G = 0.21\pm 0.02$, is close to the value $\Delta G = 0.23(6)$ determined by the NNPDF Collaboration~\cite{Nocera:2014gqa}. However, $\Delta G = 0.19(6)$ reported for $x_g \in [0.05, 1]$ by DSSV~\cite{deFlorian:2014yva} is lower than our computed value, $\Delta G = 0.28\pm 0.03$.
 Lattice QCD at physical pion mass predicts $\Delta G = 0.251(47)(16)$~\cite{Yang:2016plb}. Due to limited experimental precision, the extraction of $\Delta G$ is sensitive to the parametrization used in global analyses, leading to large uncertainties in $\Delta g(x)$, especially if a sign change of $\Delta g(x)$ is allowed at certain values of $x$~\cite{Zhou:2022wzm}. Future measurements for $x_g < 0.02$ are crucial to reduce uncertainties in $\Delta G$, a key objective for upcoming EICs~\cite{Accardi:2012qut,AbdulKhalek:2021gbh}.{ We point out, once more, that in our model, we take the gluon as the active parton and the rest of the constituents as spectator; the parameters are determined by a fit. It is nontrivial to satisfy the spin sum rule in such phenomenological spectator-type models. A discussion can be found in Ref.~\cite{Gurjar:2021dyv}.} 

	\begin{table}
		\caption{Comparison of the numerical values for the gluon spin contribution with available data at $\mu_{0} = 2$ GeV.}
        \vspace{0.15cm}
		\label{Tab:spincontributions}
		\centering
		\begin{tabular}{ c|c|c} 
			\hline\hline
            \vspace{0.05cm}
			$\Delta G^{[m,n]}=\int_{m}^{n}dx\Delta g(x)$ & Exp./global fits &  Our predictions\\
			\hline
			$\Delta G^{[0.02,\,0.3]}$	& 0.20~~~~\cite{PHENIX:2008swq}   & $0.33_{-0.027}^{+0.030}$ \\ 
			
			$\Delta G^{[0.05,\,0.2]}$	& 0.23(6)~\cite{Nocera:2014gqa}& $0.21^{+0.018}_{-0.017}$ \\ 
			
			$\Delta G^{[0.05,\,1]}$	& 0.19(6)~\cite{deFlorian:2014yva}& $0.28^{+0.031}_{-0.027}$ \\
			\hline\hline
		\end{tabular}
	\end{table}
    
\section{Gluon gravitational form factors}
The matrix elements of local operators, such as the electromagnetic current and EMT, have an exact representation in the light-front Fock state wave functions of hadrons. GFFs are obtained from the matrix elements of the EMT, $T^{\mu \nu}$, and the second moment of GPDs also yields the GFFs. For a spin-$\frac{1}{2}$ target, the standard parametrization of $T^{\mu \nu}$ with GFFs is given by~\cite{Ji:2012vj,Harindranath:2013goa},
\begin{align}\label{tensor}
\langle P',\,& \lambda'|T^{\mu \nu}_i(0)|P,\, \lambda\rangle =\overline{U}(P', \lambda')\Big[-B_i(q^2)\frac{\bar{P}^\mu\bar{P^\nu}}{M} \nonumber\\
&+(A_i(q^2)+B_i(q^2))\frac{1}{2}(\gamma^{\mu}\bar{P}^{\nu}+\gamma^{\nu}\bar{P}^{\mu}) \\
&+ C_i(q^2)\frac{q^{\mu}q^{\nu}-q^2 g^{\mu\nu}}{M}+\overline{C}_i(q^2)M g^{\mu\nu}\Big]U(P,\lambda).\nonumber
\end{align}
Here, $\bar{P}^\mu = \frac{1}{2}(P' + P)^\mu$, $q^\mu = (P' - P)^\mu$, $U(P,\lambda)$ is the Dirac spinor,  and $M$ represents the system mass. $A_i, B_i, C_i $ and  $\overline{C}_i$ are the quark or gluon GFFs. The notation $D_{i}(q^2) = 4C_{i}(q^2)$ is also used. We consider the frame where the momentum transfer $q=(0,0,\vec{q}_{\perp})$, thus $Q^2=-q^2=\vec{q}_{\perp}^{\,2}$. 

The gauge invariant symmetric form of the QCD EMT is expressed as~\cite{Harindranath:1997kk}
\begin{align}\label{emtqcd}
T^{\mu \nu} &= \frac{1}{2}\overline{\psi}\ i\left[\gamma^{\mu}D^{\nu}+\gamma^{\nu}D^{\mu}\right]\psi - F^{\mu \lambda a}F_{\lambda a}^{\nu} \nonumber\\
& + \frac{1}{4} g^{\mu \nu} \left( F_{\lambda \sigma a}\right)^2 - g^{\mu \nu} \overline{\psi} \left(
i\gamma^{\lambda}D_{\lambda} -m
\right)
\psi\,,
\end{align}
with $\psi$ and $A^{\mu}$ being the fermion and boson fields, respectively. $F^{\mu \nu}_a$ is the field strength tensor for non-Abelian gauge theory, which is expressed as
\begin{align}
F^{\mu \nu}_a = \partial^{\mu} A^{\nu}_a - \partial^{\nu} A^{\mu}_a +  g \ f^{abc} A^{\mu}_b A^\nu_c\,,
\end{align}
where the covariant derivative $
iD^{\mu} = i\overleftrightarrow{\partial}^\mu+gA^{\mu} $ such that   
$\alpha (i\overleftrightarrow{\partial}^\mu)\beta
= \frac{i}{2}\alpha\left(\partial^{\mu}\beta\right) -\frac{i}{2}\left(\partial^{\mu}\alpha\right) \beta$. In this work, we focus only on the gluonic part of the EMT given in Eq.~(\ref{emtqcd}):
	\begin{align}
	T^{\mu \nu}_g= - F^{\mu \lambda a}F_{\lambda a}^{\nu} + \frac{1}{4} g^{\mu \nu} \left( F_{\lambda \sigma a}\right)^2 .
	\end{align}
In order to calculate the GFFs, we define the matrix element of the EMT as follows
	\begin{align} \label{matrixelement}
	\mathcal{M}^{\mu \nu }_{\lambda\lambda^\prime} = \frac{1}{2}\left[\langle P'  ,\lambda'|	T^{\mu \nu }_g(0)|P,\lambda \rangle \right],
\end{align}
where  $(\lambda, \lambda^\prime) \equiv \{ \uparrow,\downarrow \}$  is the helicity of the initial and final states of the proton. The form factor $A_g(q^2)$ and $B_g(q^2)$ can  be obtained from the `good' components of EMT, $T_g^{++}$, while the GFFs $D_g(q^2)=4C_g(q^2)$ and $\overline{C}_g(q^2)$ can be extracted from the transverse component of the EMT, $T^{ij}_g$, such that $(i,j)\equiv(1,2)$. 
Using Eq.~\eqref{matrixelement}, one can derive the following relations~\cite{More:2023pcy}:
	\begin{align}\label{rhsA}
	\mathcal{M}^{++}_{\uparrow \uparrow} + \mathcal{M}^{++}_{\downarrow \downarrow} &= 2\  (P^+)^2A_g(q^2), \\
	\label{rhsB}
	\mathcal{M}^{++}_{\uparrow \downarrow} + \mathcal{M}^{++}_{\downarrow \uparrow} &= \frac{ i q_\perp^{(2)}}{M} \ (P^+)^2 B_g(q^2) .\\
		q_{\mu}\mathcal{M}^{\mu 1}_{\uparrow \downarrow} + q_{\mu}\mathcal{M}^{\mu 1}_{\downarrow \uparrow} &= -i q_\perp^{(1)}q_\perp^{(2)}M\, \overline{C}_g(q^2).
		\label{cbar3}   \\
		\mathcal{M}^{11}_{\uparrow \downarrow} +
	\mathcal{M}^{22}_{\uparrow \downarrow} &+\mathcal{M}^{11}_{\downarrow \uparrow} + \mathcal{M}^{22}_{\downarrow \uparrow} =
	i\Big[B_g(q^2)\frac{q^2}{4M}\nonumber\\
	&-D_g(q^2)\frac{3q^2}{4M}+\overline{C}_g(q^2) 2M\Big]\qp^{(2)}.\label{c3a}
	\end{align}
Using the two particle Fock states, Eq.~(\ref{state}), and the LFWFs given in Eqs.~(\ref{LFWFsuparrow}) and (\ref{LFWFsdownarrow}), we evaluate the matrix elements of $T_{g}^{++}$ and $T_{g}^{ij}$, and extract the GFFs $A_g(q^2)$, $B_g(q^2)$, $D_g(q^2)$, and $\overline{C}_g(q^2)$ from Eqs.~(\ref{rhsA})-(\ref{c3a}). The analytical expression of model results for the gluonic GFFs are as follows:
\begin{align}
    A_g(Q^2)&=2 N_g^2\int_0^1 {\rm d} x\, x^{2 b+2} (1-x)^{2 a}   \nonumber\\
&\times 
\Big(\frac{1}{1-x}\Big)^{-\frac{Q^2 (1-x)^2}{4 \kappa ^2 x^2}}
\Bigg[\Big(M-\frac{M_s}{1-x}\Big)^2 \nonumber \\
&+\frac{\Big(1+(1-x)^{2}\Big) \bigg(\frac{\kappa ^2 x^2}{\log (\frac{1}{1-x})}-\frac{1}{4} Q^2 (1-x)^2\bigg)}{(1-x)^2 x^2}\Bigg],\\
B_g(Q^2)&=-4 M N_g^2\int_0^1 {\rm d} x\, x^{2 b+1} (1-x)^{2 a+1}  \nonumber\\
&\times \Big(M-\frac{M_{s}}{1-x}\Big) \Big(\frac{1}{1-x}\Big)^{-\frac{Q^2 (1-x)^2}{4 \kappa ^2 x^2}},\\
D_g(Q^2)&=\frac{4}{3} M N_g^2 \int_0^1 {\rm d} x\, x^{2 b-1}(1-x)^{2 a+1}  \nonumber\\
& \times \Big(M-\frac{M_s}{1-x}\Big) \Big(\frac{1}{1-x}\Big)^{-\frac{Q^2 (1-x)^2}{4 \kappa ^2 x^2}},
\end{align}
\begin{align}
\overline{C}_g(Q^2)&=N_g^2 \int_0^1 {\rm d} x\, x^{2 b-1}(1-x)^{2 a}\nonumber\\
& \times \Big(M-\frac{M_s}{1-x}\Big) \Big(\frac{1}{1-x}\Big)^{-\frac{Q^2 (1-x)^2}{4 \kappa ^2 x^2}}  \nonumber\\
&\times\frac{\Big(2 \kappa ^2 x^3+Q^2 (1-x) \log (\frac{1}{1-x})\Big)}{M \log (\frac{1}{1-x})}.
\end{align}	
\begin{figure*}[htp]

\includegraphics[width=0.44\textwidth]{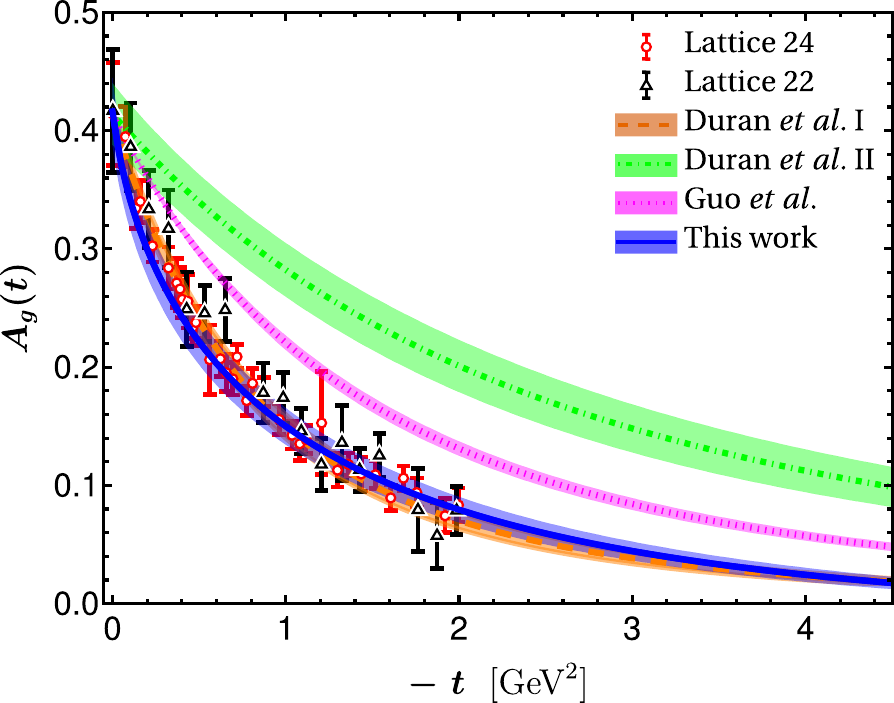}\quad\quad
\includegraphics[width=0.46\textwidth]{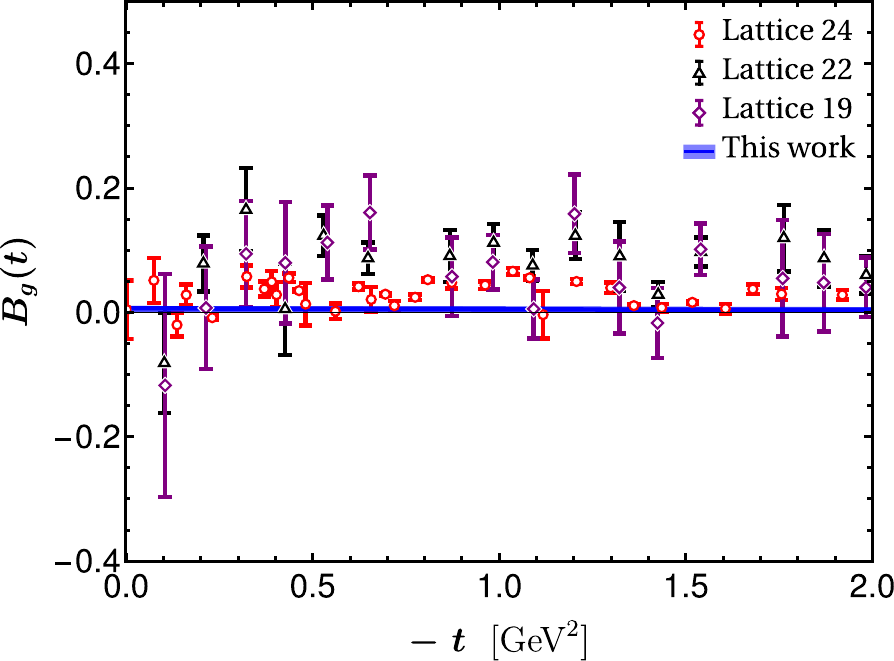}\\
\vspace{0.5cm}
\includegraphics[width=0.44\textwidth]{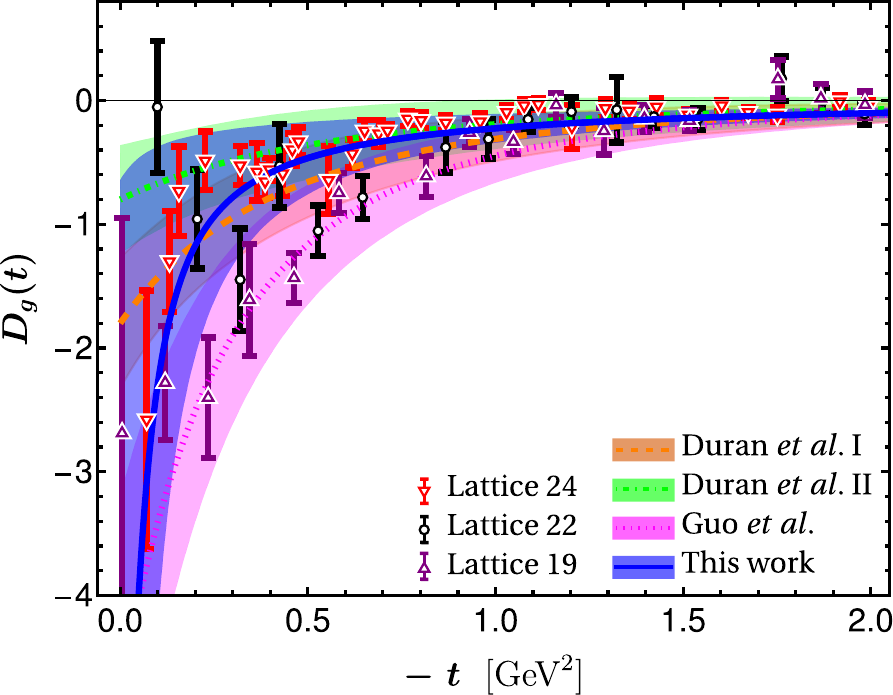}\quad\quad
\includegraphics[width=0.44\textwidth]{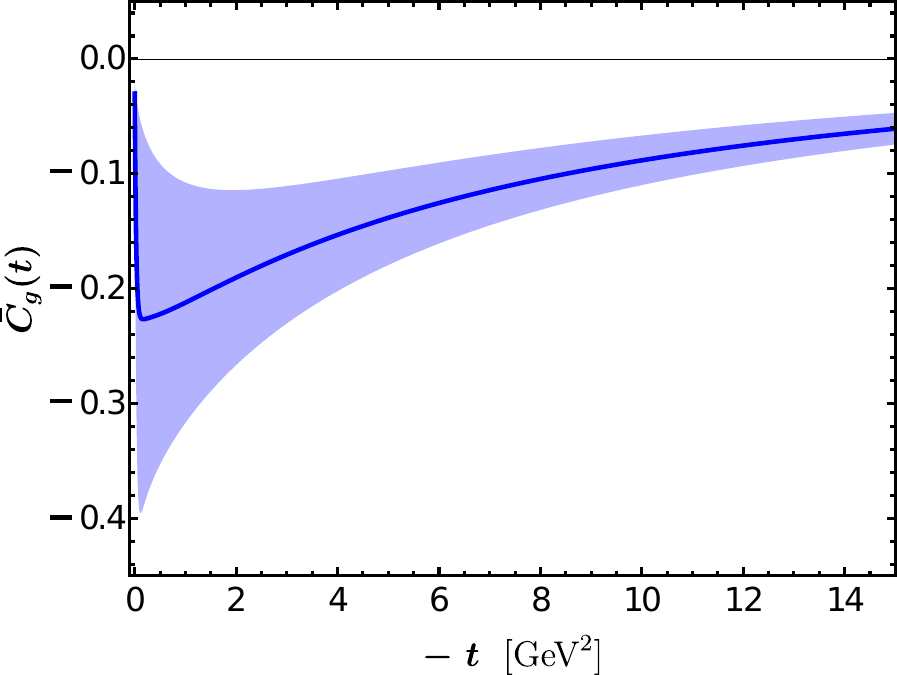}
\caption{The gluon GFFs in the proton, top-left: $A_{g}(Q^2)$, top-right: $B_{g}(Q^2)$, bottom-left $D_{g}(Q^2)$, and bottom-right $\overline{C}_{g}(Q^2)$, where $-t=Q^2$. Our results for $A_{g}(Q^2)$, $B_{g}(Q^2)$, and $D_{g}(Q^2)$ (solid blue line with blue band corresponding to the uncertainty in model parameters) are compared with the recent lattice QCD simulations~\cite{Hackett:2023rif,Pefkou:2021fni}. Our results for  $A_{g}(Q^2)$ and $D_{g}(Q^2)$ are also compared with the experimentally extracted results using `holographic QCD' inspired approach (orange band: Duran {\it et. al.} I)~\cite{Duran:2022xag} and the GPD inspired approach (green band: Duran {\it et. al.} II and magenta band: Guo {\it et. al.})~\cite{Duran:2022xag,Guo:2021ibg}.}
\label{A_B} 
\end{figure*}
The upper two panels of Fig.~\ref{A_B} display the gluon GFFs $A_g(Q^2)$ and $B_g(Q^2)$, respectively. We compare our result for $A_g(Q^2)$  with the available experimentally constrained multipole parametrization results for the gluonic GFFs presented in Refs.~\cite{Duran:2022xag,Guo:2021ibg} and the lattice QCD simulations~\cite{Hackett:2023rif,Pefkou:2021fni}. The GFF $A_{g}(Q^2)$ is found to be consistent with the latest lattice QCD predictions as well as with the `holographic QCD' inspired approach~\cite{Mamo:2022eui,Mamo:2021krl}, (I) to the analysis of experimental data in Ref.~\cite{Duran:2022xag} and disfavor the GPD inspired approach~\cite{Guo:2021ibg} (II). A more recent analysis~\cite{Guo:2023pqw} including an update to the GPD inspired analysis method, as well as additional experimental data~\cite{GlueX:2023pev}, is in less tension with our results. 
We find a reasonable agreement of the GFF $B_{g}(Q^2)$  with the recent lattice
QCD results~\cite{Hackett:2023rif,Pefkou:2021fni} showing slightly positive distribution over $Q^2$. {However, it is shown to be negative in a recent calculation for a perturbative model of a quark dressed with a gluon ~\cite{More:2023pcy}}. 
\begin{table}[htp]
\footnotesize
\caption{Comparison of the {\it Druck-term}, $D_g(0)$ between our result, lattice QCD prediction and phenomenological extractions  using different approaches.}
\label{table:D0}
\vspace{0.2cm}
\centering
\begin{tabular}{lc}
\hline \hline
Approaches &  $D_g(0) $ \\
\hline
This work & $-8.61^{+0.71}_{-0.75}$ \\
Lattice QCD~\cite{Shanahan:2018pib} & $-10(3)$ \\
Lattice QCD~\cite{Hackett:2023rif} & $-2.57(84)$  \\
Extracted~(Holo.)~\cite{Duran:2022xag}& $-1.80(52)$\\
Extracted~(GPDs)~\cite{Duran:2022xag}& $-0.80(44)$\\
Extracted~(GPDs)~\cite{Guo:2023pqw}& $-5.96(108)$\\
Dyson Schwinger Method  \cite{Yao:2024ixu} & $ -1.294(33)$ \\
\hline \hline
\end{tabular}
\end{table}
\begin{figure}[htp]
\includegraphics[width=0.44\textwidth]{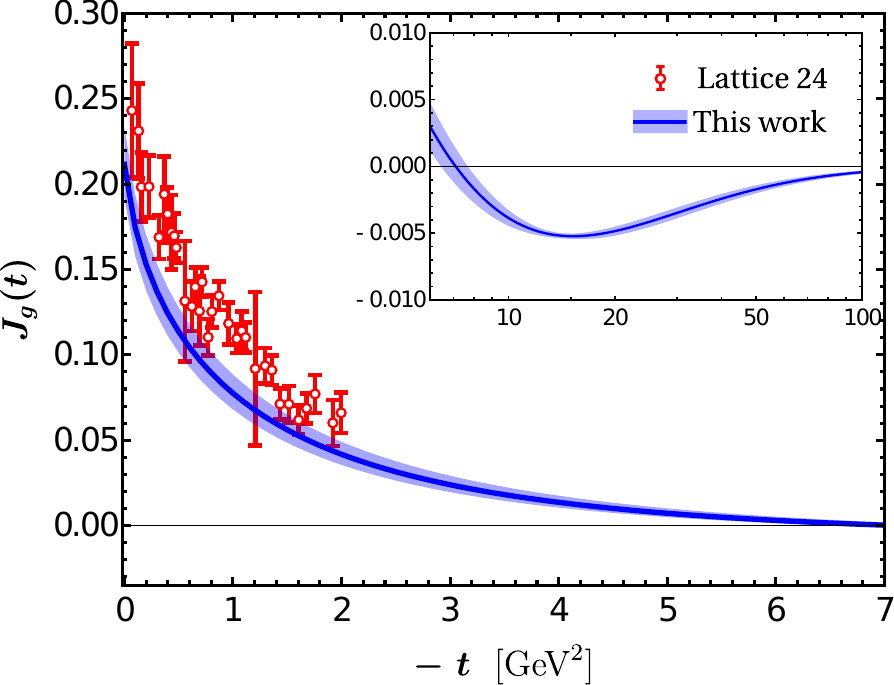}
\caption{ Our results for gluon angular momentum GFFs $J_{g}(t)$ (solid blue line with blue band corresponding to the uncertainty in model parameters) for proton compared with Lattice 2024~\cite{Hackett:2023rif} results. The inset shows the zero crossing of $J_{g} (t)$ for $-t=Q^{2}>7~\text{GeV}^{2}$.}
\label{fig:GFFJg} 
\end{figure}

\begin{figure*}[htp]
\centering
\includegraphics[width=0.44\textwidth]{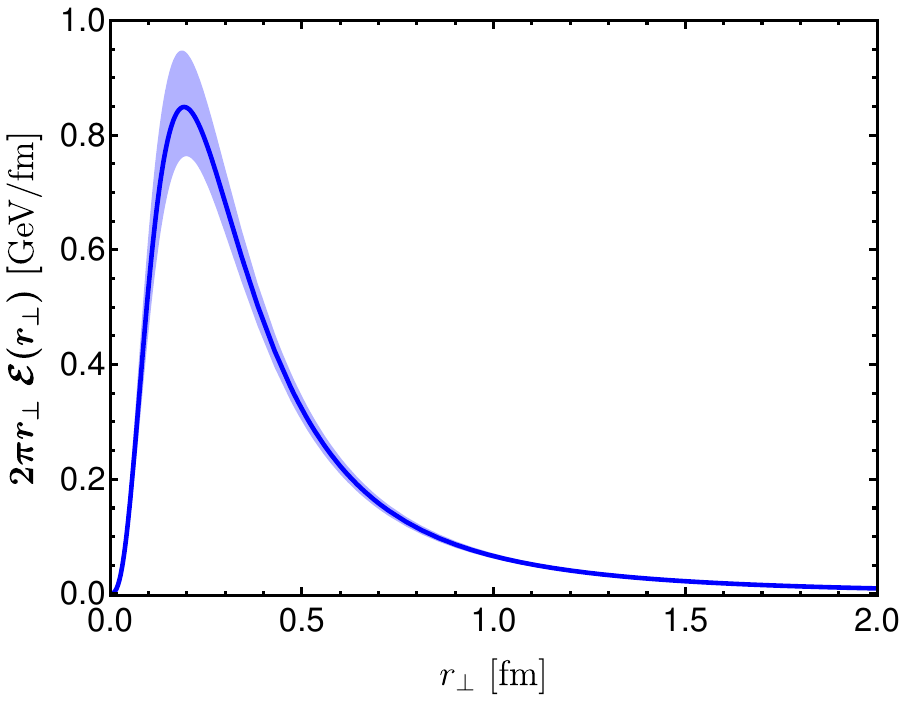}
\quad\quad
\includegraphics[width=0.44\textwidth]{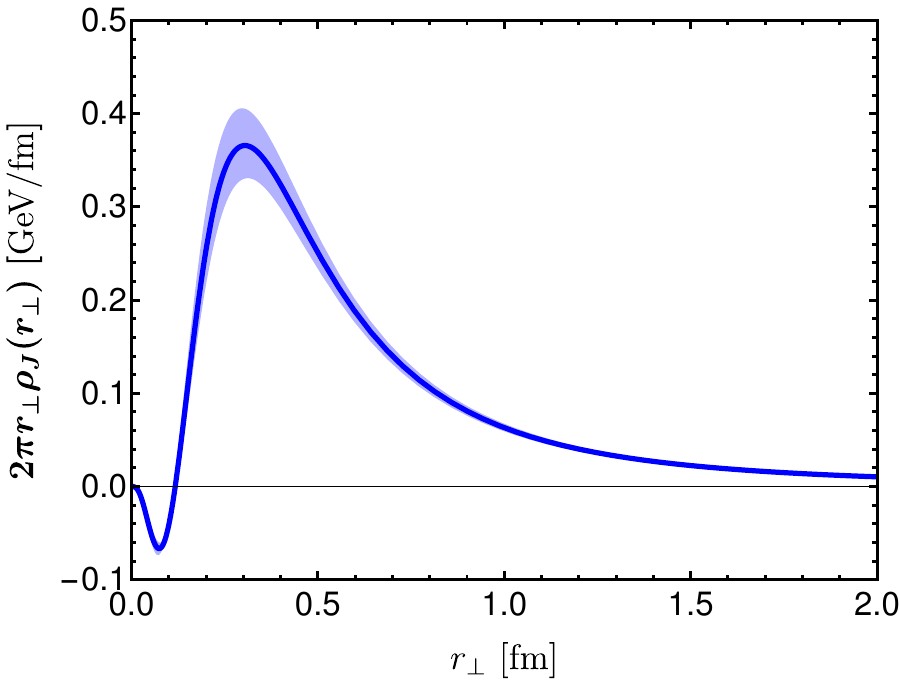}
\caption{The {light-front { mass}}  $\mathcal{E}(r_\perp)$ (left) and angular momentum $\rho_{J}(r_\perp)$ (right) distributions weighted by $2\pi r_\perp$. The band is corresponding to the uncertainty in model parameters.}
\label{density:Longi} 
\end{figure*}

The gluon GFFs $D_g(Q^2)$ and $\overline{C}_g(Q^2)$ are shown in the lower panels of Fig.~\ref{A_B}. Both the form factors are found to be  negative over $Q^2$. We compare our prediction for $D_g(Q^2)$ to lattice QCD and the experimentally extracted results using the holographic QCD and GPD approaches. We find a good agreement between our $Q^2$-dependency results and the latest lattice QCD simulations~\cite{Hackett:2023rif,Pefkou:2021fni,Shanahan:2018pib}. Our $D_{g}(Q^2)$ is also compatible with those extracted results following the holographic QCD and GPD approaches~\cite{Duran:2022xag,Guo:2021ibg,Guo:2023pqw}. The forward limit, $D_g(0)$ ({\it Druck-term}), is comparable with the lattice QCD simulation reported in Ref.~\cite{Shanahan:2018pib} but strongly differs from the recent lattice calculation~\cite{Hackett:2023rif} and extracted results using various approaches as summarized in Table~\ref{table:D0}. 
Having said that our results is close to a more recent analysis~\cite{Guo:2023pqw} including an update to the GPD inspired analysis method, as well as additional experimental data~\cite{GlueX:2023pev}. There is no available data for the $\overline{C}_g(Q^2)$ from lattice QCD or experiments. { Due to the conservation of EMT in QCD, the total $\overline{C} (Q^2)$ summing over all constituents should equal zero for all values of $Q^2$, i.e., $\sum_{i=q,g}\overline{C}_{i}(Q^2)=0$. However, individually this allows non-zero $\overline{C}_{q/g} (Q^2)$.}  In our model, focusing on the gluon, our prediction for $\overline{C}_g(Q^2)$ form factor exhibits a negative distribution having a peak at small $Q^2$-region. It has a long tail compared to the other gluonic GFFs. In the forward limit, we find that 
$\overline{C}_g(0)=-0.030 \pm 0.010$. { In a perturbative calculation for a quark dressed with a gluon}, it is also found to be negative, $\overline{C}_g(0)=-0.014$~\cite{More:2023pcy}.

{ Together $A_g(Q^2)$ and $B_g(Q^2)$ GFFs, we obtain the contribution to the proton’s total angular momentum from the gluon:
\begin{align}
    J_{g}(Q^2=0)=\frac{1}{2}\big[A_{g}(0)+B_{g}(0)\big].
\end{align}}
{Fig.~\ref{fig:GFFJg} depicts the variation of the gluon angular momentum GFF, $J_{g}(Q^2)$, with momentum transfer $Q^2$. Our model results for angular momentum (solid blue line with blue band) are compared with recent Lattice QCD results~\cite{Hackett:2023rif}. The inset plot shows the zero crossing of $J_{g}(Q^2)$ for larger values of $Q^2$, specifically, $Q^2 > 7~\text{GeV}^{2}$.} 
{ We obtain $J_{g} = 0.206 \pm 0.013$ within uncertainties in model parameters $a$ and $b$, and find a reasonable agreement with recent lattice QCD calculations: $J_g = 0.231(11)(22)$~\cite{Wang:2021vqy}, $J_g = 0.187(46)$~\cite{Alexandrou:2020sml}, and $J_g = 0.255(13)$~\cite{Hackett:2023rif}. Additionally, our result is consistent with the Bethe-Salpeter approach, which gives $J_g = 0.208 \pm 0.06$~\cite{Yao:2024ixu} }.

\begin{figure*}[htp]
\centering
\includegraphics[scale=0.52]{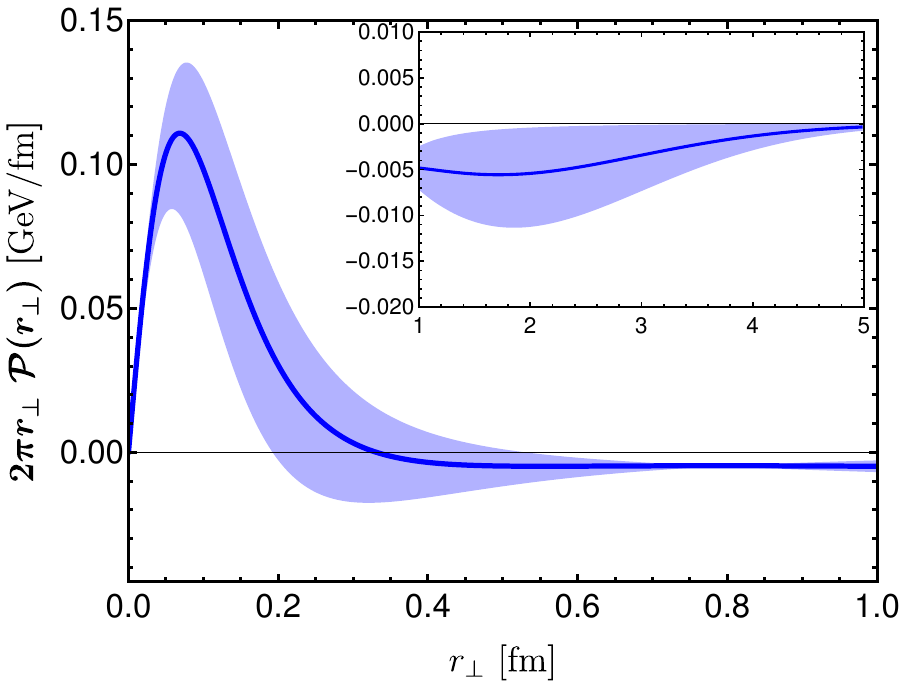}\quad\quad
\includegraphics[scale=0.52]{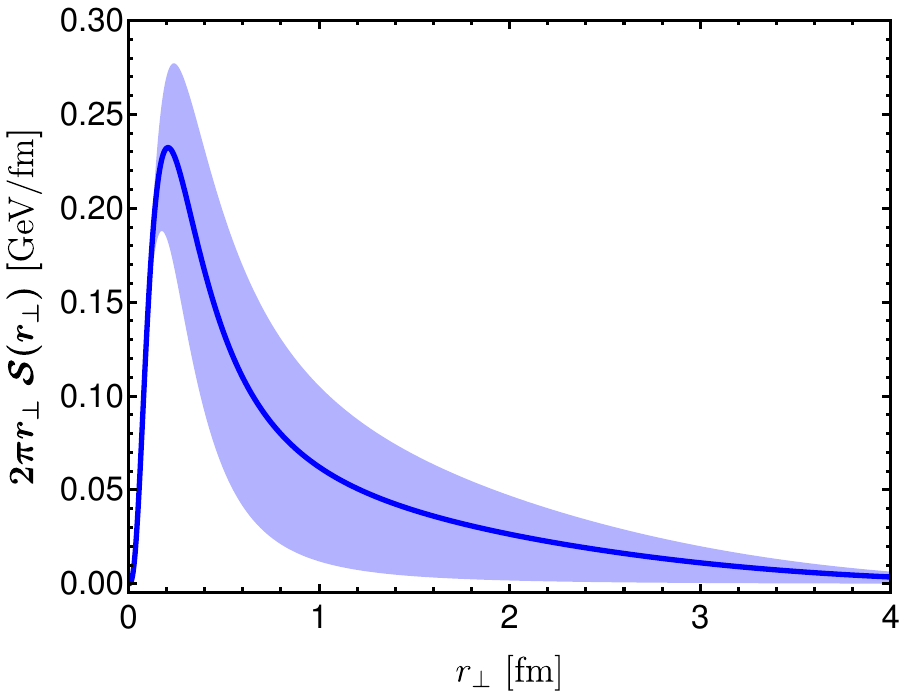}
\caption{The two-dimensional pressure $\mathcal{P}(r_\perp)$ (left) and shear-force ${S}(r_\perp)$ (right) distributions weighted by $2\pi r_\perp$. {The inset shows the large $r_{\perp}$ behavior of pressure distribution and the band is corresponding to the uncertainty in model parameters.}}
\label{density:ps} 
\end{figure*}

\subsection{Light-front densities and radii}
From the gluon GFFs $A_{g}(Q^2) $ and $ D_{g}(Q^2)$, we can compute a range of transverse densities for the system, including the light-front momentum density, energy density, pressure, and shear density, among others~\cite{Xu:2024cfa}.  
The corresponding light-front distributions in the two-dimensional (2D) transverse plane such as the { light-front} { momentum density $ \varepsilon^{({\rm LF})}(r_{\perp}) $}, angular momentum density $ \rho_{J}^{({\rm LF})}(r_{\perp}) $, pressure $ p^{({\rm LF})}(r_{\perp})$, and shear force $ s^{({\rm LF})}(r_{\perp}) $ are obtained via a 2D  Fourier transform~\cite{Freese:2021czn,Kim:2021jjf}:
\begin{align}
 \varepsilon^{(\rm LF)}(r_{\perp}) &=  P^{+} \tilde{A}(r_{\perp}), \\
\rho^{(\rm LF)}_{J}(r_{\perp}) &= -\frac{1}{2}r_{\perp} \frac{{\rm d}}{{\rm d}r_{\perp}}
  \tilde{J}(r_{\perp}),  \\
  s^{(\rm LF)}(r_{\perp}) &= -\frac{1}{4P^{+}}
   r_{\perp}\frac{1}{{\rm d}r_{\perp}} \frac{1}{r_{\perp}}
   \frac{{\rm d}}{{\rm d}r_{\perp}}  \tilde{D}(r_{\perp}), \\
   p^{(\rm LF)}(r_{\perp}) &=
   \frac{1}{8P^{+}}\frac{1}{r_{\perp}}
   \frac{{\rm d}}{{\rm d}r_{\perp}}r_{\perp}\frac{{\rm d}}{{\rm d}r_{\perp}}\tilde{D}(r_{\perp}),   
 \label{eq:2DFF}
\end{align}
where the 2D Fourier transform of the
corresponding GFFs are defined as follows 
\begin{align}
\tilde{F}(r_{\perp}) = \int \frac{{\rm d}^{2}
  \vec{q}_{\perp}}{(2\pi)^{2}}e^{-i\vec{q}_{\perp} \cdot
  \vec{r}_{\perp}} F(-\vec{q}^{\,2}_{\perp}). 
\end{align}
The vectors $\vec{r}_{\perp}$ and $\vec{q}_{\perp}$ represent the position and momentum in the two-dimensional plane transverse to the proton’s motion, respectively, and $P^{+}$ denotes the light-cone momentum. For convenience, the { light-front { mass}}, shear force, and pressure distributions are redefined by including Lorentz factors, as follows~\cite{Panteleeva:2021iip}: 
\begin{align}
\mathcal{E}(r_\perp)&= \frac{M}{P^{+}}\varepsilon^{(\rm LF)}(r_\perp), \quad
  {S}(r_\perp)= \frac{P^{+}}{2M}s^{(\rm LF)}(r_\perp),  \nonumber\\
  \mathcal{P}(r_\perp)&=\frac{P^{+}}{2M}p^{(\rm LF)}(r_\perp) .
\label{eq:2DFFcov}
\end{align}
 Note that the longitudinal boost does not affect the longitudinal component of the angular momentum, its distribution does not require an additional Lorentz factor~\cite{Lorce:2017wkb}. {It is also essential to clarify that $\varepsilon^{(\rm LF)}(r_\perp)$ represents the 2D light-front momentum density, whereas $\mathcal{E}(r_\perp)$ corresponds to the 2D light-front mass density in Eq.~\eqref{eq:2DFFcov}.}

The pressure and shear distributions { are expected to satisfy } the von Laue stability condition~\cite{Laue:1911lrk,Polyakov:2018zvc,Freese:2021czn}:
\begin{align}
  \int \mathrm{d}^2\vec{r}_\perp \,
   \mathcal{P}(r_\perp)
  =
  0\label{vl}
  \,,
  \\
      \frac{\mathrm{d}\mathcal{P}(r_\perp)}{\mathrm{d}r_\perp} + \frac{1}{2} \frac{\mathrm{d}{S}(r_\perp)}{\mathrm{d}r_\perp} + \frac{1}{r_\perp} {S}(r_\perp) = 0
  \,,
\end{align}
 which are light-front analogy in the transverse plane to Eq.~(30) of Ref.~\cite{Polyakov:2018zvc}.
One can also obtain expressions for the radial and tangential forces within a hadron~\cite{Lorce:2018egm,Freese:2021qtb}:
\begin{align}
  \mathcal{P}_r(r_\perp)
  &=
    \mathcal{P}(r_\perp) + \frac{1}{2} {S}(r_\perp) ,
  \\
  \mathcal{P}_t(r_\perp)
  &=
    \mathcal{P}(r_\perp) - \frac{1}{2} {S}(r_\perp)
  \,.
\end{align}
Note that in Refs.~\cite{Polyakov:2018zvc,Freese:2021czn}, $\mathcal{P}_r(r_\perp)$ is called a ``normal force,'' but we avoid this term here to emphasize that the net force across the hadron is zero. Additionally, Refs.~\cite{Polyakov:2018zvc, Lorce:2018egm, Freese:2021czn} suggest $\mathcal{P}_r(r_\perp) \geq 0$ as a stability condition, although there are no sign constraints on $\mathcal{P}_t(r_\perp)$.

\begin{figure*}[htp]
\centering
\includegraphics[width=0.44\textwidth]{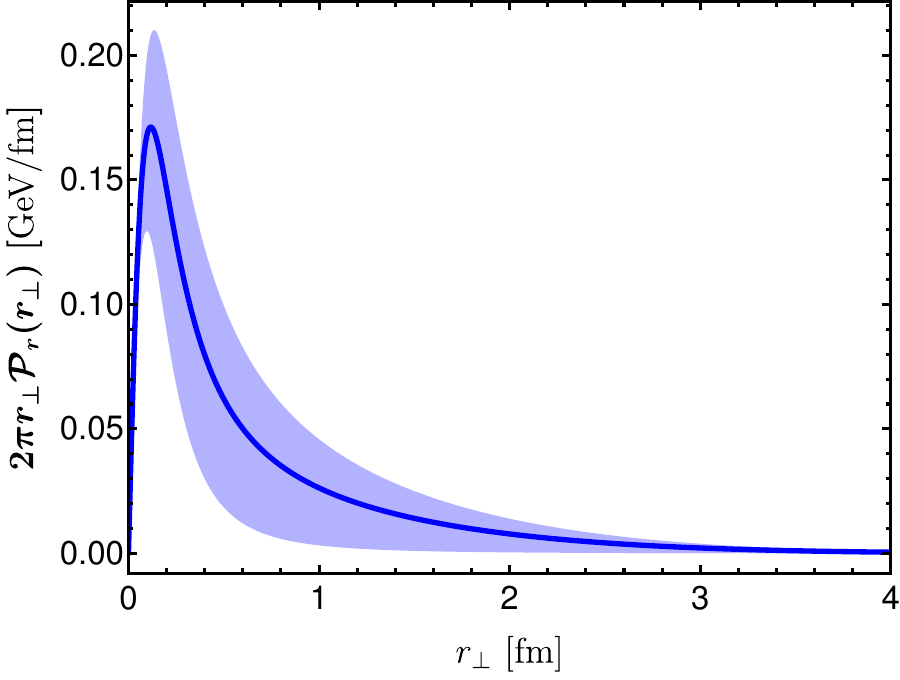}
\quad\quad
\includegraphics[width=0.44\textwidth]{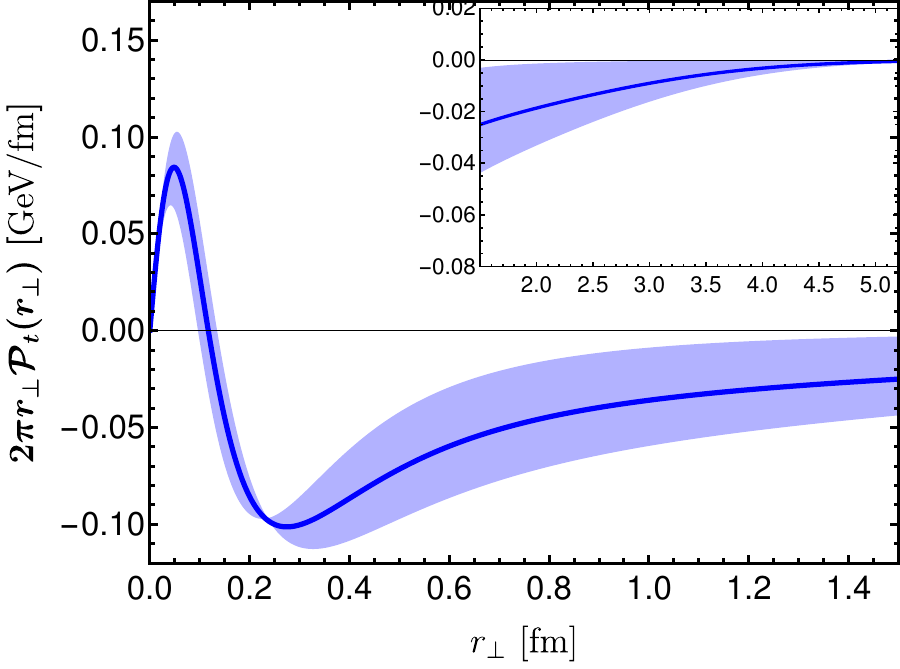}
\caption{The two-dimensional radial, $\mathcal{P}_r(r_\perp)$ (left) and tangential, $\mathcal{P}_t(r_\perp)$ (right) force distributions weighted by $2\pi r_\perp$. { The inset shows the large $r_{\perp}$ behavior of tangential pressure and the band is corresponding to the uncertainty in model parameters.}}
\label{density:prpt}
\end{figure*}

 \begin{figure}[htp]
    \centering
    \includegraphics[width=0.44\textwidth]{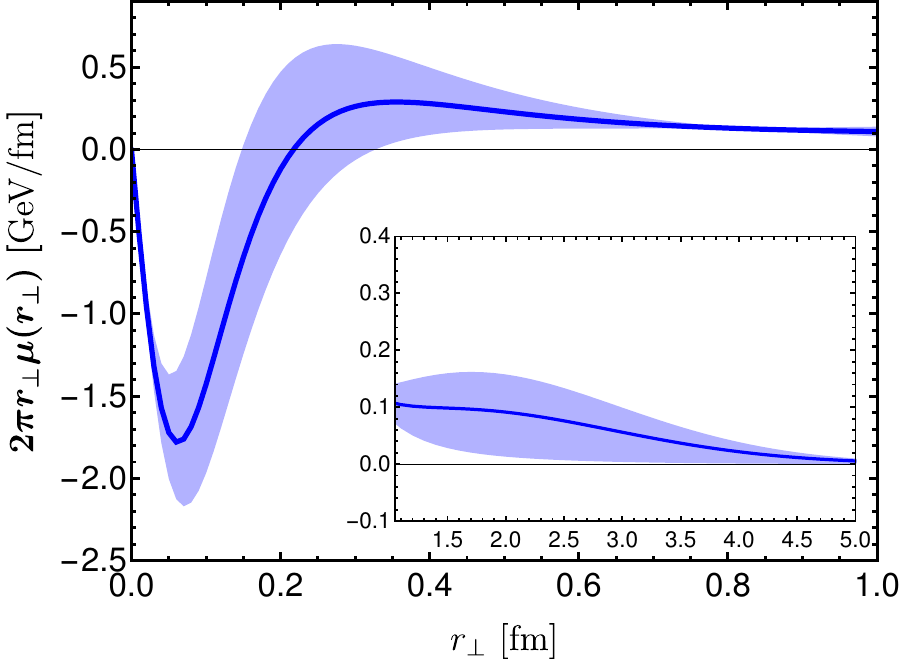}
    \caption{ The two-dimensional Galilean energy distributions weighted by $2\pi r_\perp$. The inset shows the large $r_{\perp}$ behavior of the Galilean energy and the band is corresponding to the uncertainty in model parameters.}
    \label{fig:2D_Galilean_energy}
\end{figure}
Fig.~\ref{density:Longi} shows the 2D light-front distributions for the { mass} density, $\mathcal{E}(r_\perp)$, and the { gluon} angular momentum density, $\mathcal{\rho}_{J}(r_\perp)$, each weighted by $2\pi r_\perp$. { The light-front { mass}  distribution}  peaks around  $r_\perp \sim 0.2$ fm. It  decreases monotonically as the transverse distance from the center of the proton, $r_\perp = 0$~fm, increases.  {  The angular momentum distribution for the gluon show a slightly different qualitative nature, it is negative near the core and then become positive at larger distance.  Also it shows a slightly broader profile, peaking around $r_\perp \sim 0.3$~ fm.}  { This behavior arises because the angular momentum GFF $J_g(Q^2)$ is negative at large momentum transfer, specifically for, $Q^2>7~\text{GeV}^{2}$ as shown in the inset of Fig.~\ref{fig:GFFJg}.} 
{ A similar qualitative behavior is observed for the { mass}  distribution of the quarks in a similar light front spectator model, as discussed in Refs.~\cite{Chakrabarti:2020kdc,Choudhary:2022den}, whereas the angular momentum distribution for quarks is positive even near the core of the nucleon. Quantitatively, the angular momentum density of quarks and gluons are comparable in such models, whereas the light-front { mass}  density is significantly lower for gluons compared to quarks~\cite{Choudhary:2022den}.}

Fig.~\ref{density:ps} illustrates the gluonic pressure and shear distributions, $\mathcal{P}(r_\perp)$ and ${S}(r_\perp)$, each weighted by $2\pi r_\perp$, highlighting key features of the pressure distribution. The pressure profile displays a positive density core at the center (where $r_\perp = 0$), surrounded by a ring of negative density that extends outward with a long tail. This behavior is in accordance with the von Laue { stability} condition, Eq.~\eqref{vl}, which ensures the balance of internal forces in a stable system~\cite{Polyakov:2018zvc,Polyakov:2002yz}. { For the pressure distribution coming from quarks, a similar behaviour is seen in a spectator type model \cite{Chakrabarti:2020kdc}}. According to the von Laue condition, the pressure distribution must change sign at some point. Our results exhibit a positive core region followed by a surrounding negative region, indicating the presence of both outward and inward forces within the proton. { The pressure distribution { is positive}  towards the center, peaking at a very small transverse separation of $r_\perp \approx 0.1$ fm. It changes sign around $r_\perp \sim 0.33$ fm and vanishes at large $r_\perp$, as shown in the inset plot. In \cite{Pefkou:2021fni}, the distributions are evaluated using both the tripole fit and the $z$-expansion fit. Our results show similar trend to the tripole fit in the infinite momentum frame. In contrast, the gluon pressure distribution in \cite{Shanahan:2018nnv} is more wide around the peak which is about $r \approx 0.5$ fm and changing sign at approximately $r_\perp \approx 1$ fm while ref. \cite{Yao:2024ixu} results peaks around $r \approx 0.2$ fm and flip signs on $r \approx 0.5$ fm. However, the results in \cite{Yao:2024ixu,Shanahan:2018nnv} are obtained in 3D Breit frame.  }

The shear distribution is linked to properties such as surface tension and surface energy, which are generally positive in stable hydrostatic systems~\cite{Polyakov:2018zvc}. {Our result for the shear distribution is shifted more towards the center and peaks at small value of $ r_\perp \approx 0.2$ fm while in Ref. \cite{Yao:2024ixu,Pefkou:2021fni} is peaked around $ r_\perp \approx 0.5$ fm}. Consistent with previous studies, our results confirm that the shear distribution remains positive for all values of $r_\perp$, thereby satisfying the local stability condition ${S}(r_\perp) > 0$. { However in Breit frame \cite{Shanahan:2018nnv}, it is noticed that the shear distribution negative close to the center up to $r \approx 0.5 $ fm. }

Fig.~\ref{density:prpt} presents our results for the radial and tangential forces associated with the gluonic contribution. A key observation is the consistently positive nature of $ \mathcal{P}_r(r_\perp) $, which peaks around $ r_\perp \sim 0.15 $ fm, satisfying the local stability condition $ \mathcal{P}_r(r_\perp) > 0 $. In contrast, $ \mathcal{P}_t(r_\perp) $ exhibits a dual character: a positive core, indicating repulsive forces, followed by a negative region that signifies attractive forces, with a zero-crossing around $ r_\perp \sim 0.12 $ fm. The peak of the repulsive force appears near $ r_\perp \sim 0.05 $ fm, while the maximum attractive force, crucial for binding, is stronger and located around $ r_\perp \sim 0.26 $ fm. Notably, this binding force has a larger magnitude than the repulsive component.

The $r_\perp^2$-weighted integral of gluonic {mass} density $\mathcal{E}(r_\perp)$ defines the square of a light-front {mass} radius of the gluon, which is relevant to the
3D radii in the Breit frame as~\cite{Freese:2021czn,Kim:2021jjf}:
\begin{align}
  \label{eqn:radius:p+}
  \langle r_{\perp, \mathrm{mass}}^2 \rangle_{g}
  &\equiv
  \frac{\int \mathrm{d}^2\vec{r}_\perp \,
  r_\perp^2
 \mathcal{E}(r_\perp)}{\int \mathrm{d}^2\vec{r}_\perp \,
 \mathcal{E}(r_\perp)}= 
  -\frac{4}{A_g(0)}
  \frac{\mathrm{d}A_g(Q^2)}{\mathrm{d}Q^2}\bigg|_{Q^2=0}
  \nonumber\\ &=\frac{2}{3}\langle r_{\mathrm{mass}}^2 \rangle_{g}+\frac{D_g(0)}{M^2}  \,.
\end{align}
Thus, the light front formalism provides a physical interpretation for the
slope of $A_g(Q^2)$.
Meanwhile, the 2D mechanical radius of gluon can be  defined as~\cite{Freese:2021czn,Kim:2021jjf}
\begin{align}
  \label{eqn:radius:mechanical}
  \langle r_{\perp, \mathrm{mech}}^2 \rangle_{g}
  &=
  \frac{
    \int \mathrm{d}^2 \vec{r}_\perp \,
    {r}_\perp^2
    \mathcal{P}_r(r_\perp)
  }{
    \int \mathrm{d}^2 \vec{r}_\perp \,
    \mathcal{P}_r(r_\perp)
  }\nonumber=
  \frac{
    4D_g(0)
  }{
    \int_{-\infty}^0 \mathrm{d}t \, D_g(t=-Q^2)
  }\\
  &=\frac{2}{3}\langle r_{\mathrm{mech}}^2 \rangle_{g}
  \,,
\end{align}
which is the light front analogue of
Ref.~\cite{Polyakov:2018zvc}'s Eq.~(41). 

\begin{table}[hbt!]
\footnotesize
\caption{Comparison of the gluonic mass and mechanical radii between our results, lattice QCD predictions and phenomenological extractions using different approaches.}
\label{tab:radii}
\vspace{0.2cm}
\centering
\begin{tabular}{lcc}
\hline \hline
Approaches &  $\sqrt{\langle r_{\rm mass}^2\rangle_g} $ [fm] & $\sqrt{\langle r_{\rm mech}^2\rangle_g} $ [fm] \\
\hline
This work & $1.10^{+0.39}_{-0.55}$ & $1.44^{+0.06}_{-0.35}$\\
Lattice QCD~\cite{Hackett:2023rif} & $0.81\pm 0.07$ & $0.89\pm 0.11$ \\
Dyson Schwinger Method  \cite{Yao:2024ixu} & $ 0.46 $ & 0.42 \\
Lattice QCD~\cite{Pefkou:2021fni} & $0.74 \pm 0.02$ & -- \\
Extracted~(Holo.)~\cite{Duran:2022xag}& $0.75 \pm 0.03$ & --\\
Extracted~(GPDs)~\cite{Duran:2022xag}& $0.47 \pm 0.04$ & --\\
Extracted~(GPDs)~\cite{Guo:2023pqw}& $ 0.97 \pm 0.11$ & --\\
\hline \hline
\end{tabular}
\end{table}
{ We { obtain} the gluonic contributions to the proton’s mass and mechanical radii} 
 as $\sqrt{\langle r_{\perp, \mathrm{mass}}^2 \rangle_{g}} = 0.65^{+0.01}_{-0.01}$ fm and $\sqrt{\langle r_{\perp, \mathrm{mech}}^2 \rangle_{g}} = 0.96^{+0.04}_{-0.24}$ fm, respectively. The corresponding 3D radii in the Breit frame are shown in Table~\ref{tab:radii}. We compare our predictions with results from lattice QCD and phenomenological extractions. For the gluon mass radius, our result is larger than that predicted by the GPD and `holographic QCD'-inspired approaches to analyzing experimental data in Ref.~\cite{Duran:2022xag}, but it aligns closely with the updated analysis using the GPD approach in Ref.~\cite{Guo:2023pqw}, considering our model uncertainties. Our result for the gluon mechanical radius is larger than the value reported in recent lattice QCD simulations~\cite{Hackett:2023rif}. 
We note that the proton’s gluonic mass and mechanical radii are notably larger than its charge radius, $\sqrt{\langle r^2\rangle_{\rm c}} = 0.840^{+0.003}_{-0.002}$ fm~\cite{ParticleDataGroup:2022pth}.
 \section{Galilean energy density and pressure distributions}\label{sec_Galilean}
The two-dimensional Galilean energy density, $\mu(r_\perp)$, along with the radial pressure $\sigma_{r}(r_\perp)$, tangential pressure $\sigma_{t}(r_\perp)$, isotropic pressure $\sigma(r_\perp)$, and pressure anisotropy $\Pi(r_\perp)$, are defined as~\cite{Lorce:2018egm}:
     \begin{align} 	\label{Genergy}
 \mu(r_\perp) =& M \Bigg[ \frac{\tilde{A}(r_\perp)}{2} + \tilde{\overline{C}}(r_\perp) + \frac{1}{4M^2}\frac{1}{r_\perp}\frac{{\rm d}}{{\rm d} r_\perp} \Big( r_\perp \frac{{\rm d}}{{\rm d} r_\perp}\Big[ \nonumber\\ 
 &\frac{\tilde{B}(r_\perp)}{2} - 4\tilde{C}(r_\perp) \Big]\Big) \Bigg] ,\\
 \label{radialP}
 \sigma_{r}(r_\perp)  =& M  \left[ -\tilde{\overline{C}}(r_\perp) + \frac{1}{M^2} \frac{1}{r_\perp} \frac{{\rm d} \tilde{C}(r_\perp)}{{\rm d}r_\perp} \right] , \\
 \label{tangentialP}
 \sigma_{t}(r_\perp) =& M  \left[ -\tilde{\overline{C}}(r_\perp) + \frac{1}{M^2} \frac{{\rm d}^2\tilde{C}(r_\perp)}{{\rm d}r_\perp^2} \right], \\
 \label{totalP}
 \sigma(r_\perp) =& M  \left[ -\tilde{\overline{C}}(r_\perp) + \frac{1}{2}\frac{1}{M^2} \frac{1}{r_\perp} \frac{{\rm d}}{{\rm d}r_\perp}\left(r_\perp \frac{{\rm d}  \tilde{C}(r_\perp)}{{\rm d} r_\perp} \right) \right], \\
 \label{shearlike}
 \Pi(r_\perp)  = & M  \left[ -\frac{1}{M^2} r_\perp \frac{{\rm d}}{{\rm d}r_\perp}\left(\frac{1}{r_\perp} \frac{{\rm d} \tilde{C}(r_\perp)}{{\rm d} r_\perp} \right) \right].
 \end{align}   

The isotropic pressure and pressure anisotropy can also be determined from the radial and tangential pressures using the following relations:
 \begin{align}
     \sigma(r_\perp)&=\frac{\sigma_{r}(r_\perp)+\sigma_{t}(r_\perp)}{2}\,,\nonumber\\\Pi(r_\perp)&=\sigma_{r}(r_\perp)-\sigma_{t}(r_\perp).
 \end{align}
 
 \begin{figure*}[htp]
\centering
\includegraphics[width=0.44\textwidth]{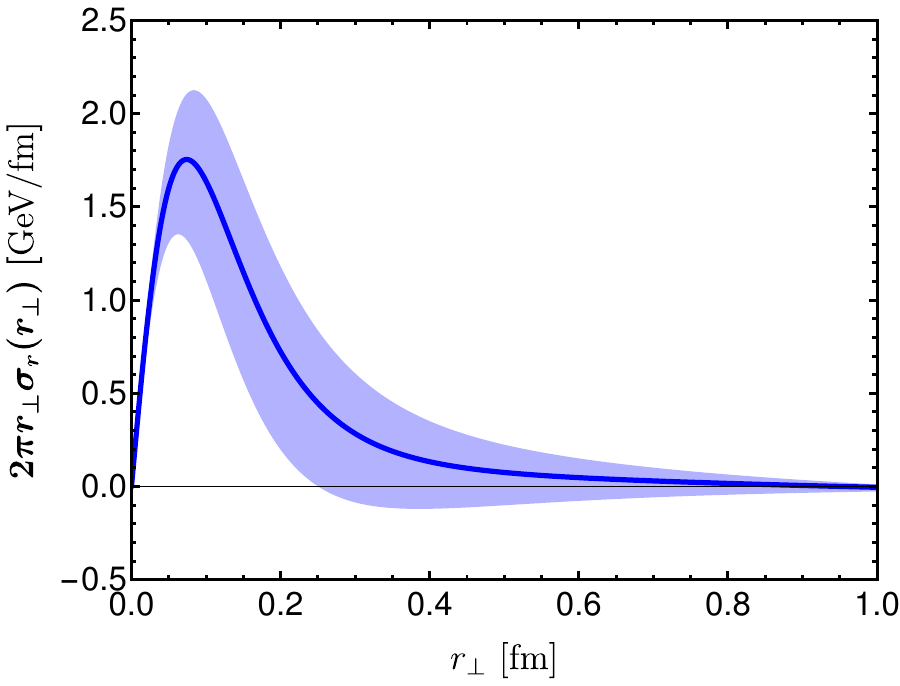}\quad\quad
\includegraphics[width=0.44\textwidth]{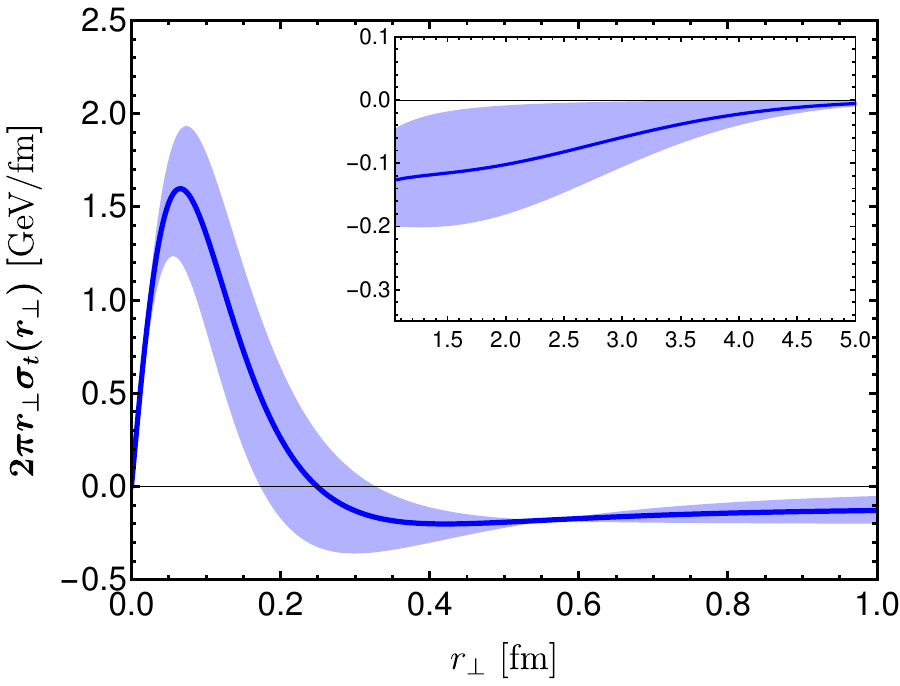}
\caption{The two-dimensional radial pressure $\sigma_{r}$ (left) and tangential pressure $\sigma_{t}$ (right) distributions weighted by $2\pi r_{\perp}$. The inset shows the large $r_{\perp}$ behavior of tangential pressure distribution and the band is corresponding to the uncertainty in model parameters.}
\label{fig:radial_tangential_pressures}
\end{figure*}

\begin{figure*}[htp]
\centering
\includegraphics[width=0.44\textwidth]{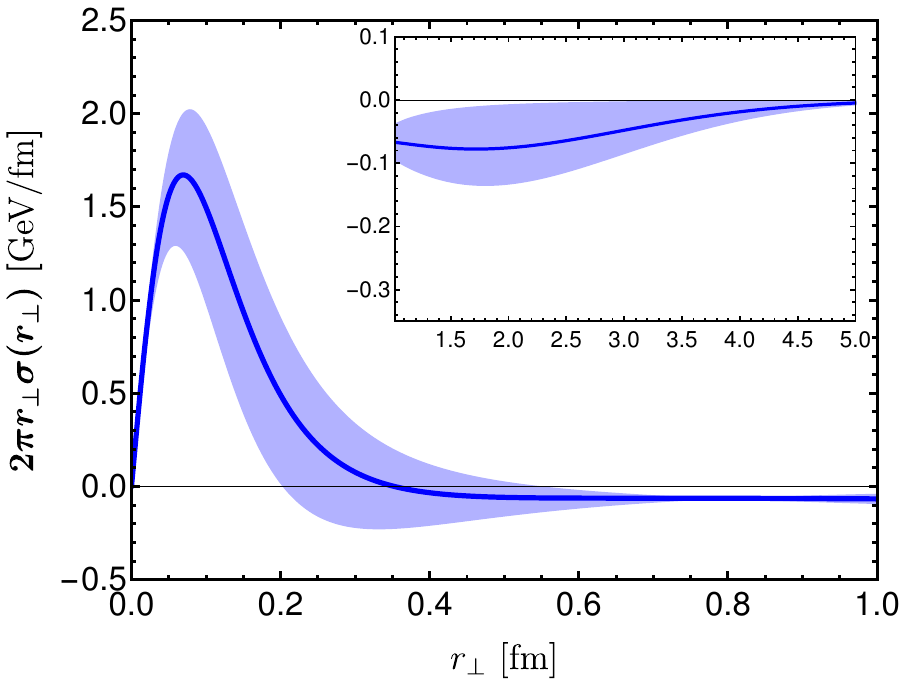}\quad\quad
\includegraphics[width=0.44\textwidth]{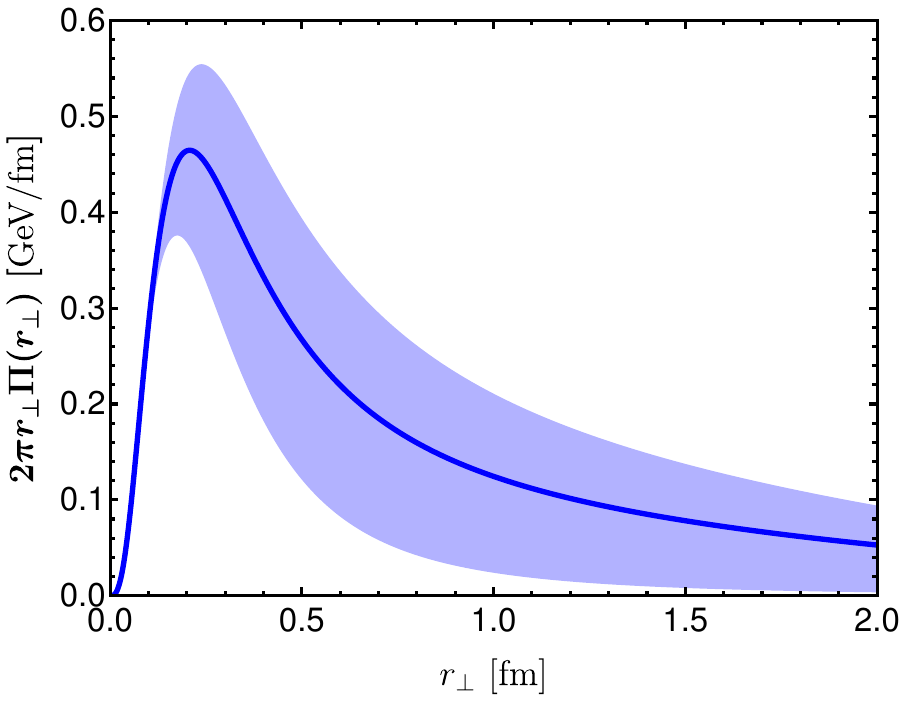}
\caption{The two-dimensional isotropic pressure $\sigma$ (left) and pressure anisotropy $\Pi$ (right) distributions weighted by $2\pi r_{\perp}$. The inset shows the large $r_{\perp}$ behavior of isotropic pressure distribution and the band is corresponding to the uncertainty in model parameters.}
\label{fig:isotropic_preeure_pressure_anisotropy}
\end{figure*}
Fig.~\ref{fig:2D_Galilean_energy} illustrates the distribution of the 2D Galilean energy, $\mu(r_\perp)$, of the gluon in proton. As seen from Eq.~\eqref{Genergy}, 
the energy density includes contributions from all four gluon GFFs. Interestingly, we observe a sharp negative peak in the 2D energy density at the center of the proton, surrounded by a ring of positive energy density. The energy density changes sign from negative to positive around $r_{\perp} \approx 0.25$ fm, attains a positive peak, and then decreases monotonically. Note that the behavior of the gluon energy density in our model exhibits the opposite trend observed in the dressed quark model~\cite{More:2023pcy}.  { The 2D Galilean energy for the quark in spectator model is also observed positive in the entire range of $r_\perp $ as discussed in \cite{Chakrabarti:2020kdc}.} 

In Fig.~\ref{fig:radial_tangential_pressures}, we present the 2D distribution of the radial pressure (left panel) and tangential pressure (right panel). The radial pressure is mostly positive over the entire range, exhibiting a repulsive nature, though the error bars extend slightly into the negative region around $r_{\perp} \approx 0.25 - 1$ fm.  
In contrast, the tangential pressure remains positive up to $r_{\perp} \approx 0.25$ fm, then changes sign and exhibits a negative distribution that extends outward over a long distance, reflecting an attractive nature. { While similar trend is observed for the quarks in \cite{Chakrabarti:2020kdc} but quark densities are peaked around $r_\perp \approx 0.5$ fm while gluon densities are inclined more towards the center of the proton.}
In the left panel of Fig.~\ref{fig:isotropic_preeure_pressure_anisotropy}, we show the isotropic pressure distribution, which is essentially the average of the radial and tangential pressures. It is positive at small $r_{\perp}$, reaching a peak around $r_{\perp}\approx0.07$ fm, then changes sign at $r_{\perp}\approx0.35$ fm, and stays negative at larger $r_{\perp}$. The right panel illustrates the pressure anisotropy, which is zero at the proton's center and peaks around $r_{\perp}\approx0.2$ fm, consistent with spherical symmetry. It remains positive across the entire impact parameter range, indicating that the radial pressure always exceeds the tangential pressure.
\section{Conclusions}
In this work, we have computed the gluon gravitational form factor (GFF) of the proton within a light-front spectator model, using light-front wave functions (LFWFs) modeled from the soft-wall AdS/QCD prediction for two-body bound states. In this simplified model, the proton is treated as a system of a struck gluon and a spin-$\frac{1}{2}$ spectator. Model parameters were determined by fitting the unpolarized gluon PDF, \( f_1^g(x) \), to the NNPDF3.0nlo global analysis. These LFWFs then provided the foundation for deriving the gluon helicity PDF and GFFs, with both calculated as model predictions.

The GFFs \( A_{g}(Q^2) \) and \( B_{g}(Q^2) \) were extracted from the \( T_{g}^{++} \) component, while \( D_{g}(Q^2) \) and \( \bar{C}_{g}(Q^2) \) were derived from the shear stress component \( T_{g}^{ij} \) (where \( i, j = 1, 2 \)). Our result for the GFF \( D_g(Q^2) \) aligns with recent lattice QCD simulations~\cite{Hackett:2023rif,Pefkou:2021fni} and experimental results from Refs.~\cite{Duran:2022xag,Guo:2021ibg}. For \( A_g(Q^2) \), our result is consistent with lattice QCD findings, though additional constraints provided by comparisons with the experimental results in Ref.~\cite{Duran:2022xag} highlight distinctions between different experimental data analyses. Notably, our result for \( A_g(Q^2) \) is consistent with the `holographic QCD'-inspired approach~\cite{Mamo:2022eui,Mamo:2021krl} (Duran~{\it et.~al.~}I) to the analysis in Ref.~\cite{Duran:2022xag} and disfavors the GPD-inspired approach~\cite{Guo:2021ibg} (Duran~{\it et.~al.~}II). A more recent analysis~\cite{Guo:2023pqw}, which updates the GPD-inspired method and includes additional experimental data~\cite{GlueX:2023pev}, shows reduced tension with our results.

Using the GFFs in the gluon-spectator model, we analyzed the light-front gluon distributions within the proton, including the pressure, $\mathcal{P}(r_\perp)$, and shear force, ${S}(r_\perp)$. We observed a positive core and a long negative tail for $\mathcal{P}(r_\perp)$, while ${S}(r_\perp)$ remained consistently positive, in alignment with other theoretical approaches. Additionally, the radial pressure, $\mathcal{P}_r(r_\perp)$, is uniformly repulsive, whereas the tangential pressure, $\mathcal{P}_t(r_\perp)$, exhibits repulsion at the center and attraction toward the periphery. These results are generally consistent with experimental and theoretical expectations. We have computed the proton's gluonic mass and mechanical radii, finding that the gluonic mass radius is larger than the charge radius. Our finding contrasts with the phenomenological model-based experimental extraction reported in Ref.~\cite{Duran:2022xag}, but aligns with a more recent analysis incorporating additional experimental data~\cite{Guo:2023pqw}. 
Our result for the gluonic mechanical radius is also larger than the proton’s charge radius and is roughly consistent with recent lattice QCD simulations~\cite{Hackett:2023rif}, when considering uncertainties. { We have also evaluated the results for the two-dimensional Galilean energy density, radial pressure, tangential pressure, isotropic pressure, and pressure anisotropy for the gluon in the model. }

\begin{acknowledgments}
The authors acknowledge fruitful discussions with  X.~Zhao, Y.~Li, S.~Xu, S.~Nair, S.~Kaur and S.~Saha. CM is supported by new faculty start up funding by the Institute of Modern Physics, Chinese Academy of Sciences, Grant No. E129952YR0.  CM also thanks the Chinese Academy of Sciences Presidents International Fellowship Initiative for the support via Grants No. 2021PM0023. A. M. would like to thank ANRF MATRICS (MTR/2021/000103) for funding.
\end{acknowledgments}

\appendix
\section{Overlap representation of matrix elements of $T_{g}^{\mu\nu}$ in terms of LFWFs}
\subsection{Matrix elements of $T_{g}^{++}$}
The nucleon spin non-flip and spin-flip gluon gravitational form factors, $A_{g}(q^{2})$ and $B_{g}(q^{2})$ in Eqs.~\eqref{rhsA} and \eqref{rhsB}, are derived from the matrix elements of $T^{++}$. In the overlap representation of LFWFs, these matrix elements are expressed as:
\begin{widetext}
\begin{align}
	\langle P^{\prime},\lambda^{\prime}|T^{++}_{g}|P,\lambda \rangle = 2(P^+)^2   \int\frac{{\rm d}^2\bfk {\rm d}x}{16\pi^3}  \, 
	 x\,\sum_{\lambda_{s}}\bigg[\psi_{+1, \lambda_{s}}^{\lambda^{\prime} *}(x,\bfk^{\prime})\psi_{+1,\lambda_{s}}^{\lambda}(x,\bfk) +\psi_{-1, \lambda_{s}}^{\lambda^{\prime} *}(x,\bfk^{\prime})\psi_{-1,\lambda_{s}}^{\lambda}(x,\bfk)\bigg].
\end{align}
where $\bfk^{\prime}=\bfk+(1-x)\bfq$. 
\end{widetext}
%
	\subsection{Matrix elements of $T_{g}^{ij}$}
The gravitational form factors $\overline{C}_{g}(q^2)$ and $D_{g}(q^2)$ are derived from the matrix elements of the transverse components of the energy-momentum tensor (EMT), $T_{g}^{ij}$, as given in Eqs.~\eqref{cbar3} and \eqref{c3a}. We evaluate $\overline{C}_{g}(q^2)$ using EMT conservation equation as follows:
		\begin{align}
	q_{\mu}\mathcal{M}^{\mu 1}_{\uparrow \downarrow} + q_{\mu}\mathcal{M}^{\mu 1}_{\downarrow \uparrow} &= -i \qp^{(1)}\qp^{(2)}M\, \overline{C}_g(q^2).
	\end{align}
			This calculation requires the matrix elements of  $ T_{g}^{11} $, $ T_{g}^{21} $, and $ T_{g}^{+1} $,  
    \begin{align}
              &-i \qp^{(1)}\qp^{(2)}M\, \overline{C}_g(q^2)  =\frac{(\bfq)^{2}}{2P^{+}} \left(\mathcal{M}^{+1}_{\uparrow\downarrow}+\mathcal{M}^{+1}_{\downarrow\uparrow}\right)
              \nonumber\\
              &-\frac{\qp^{(1)}}{2} \left(\mathcal{M}^{11}_{\uparrow\downarrow}+\mathcal{M}^{11}_{\downarrow\uparrow}\right)-\frac{\qp^{(2)}}{2} \left(\mathcal{M}^{21}_{\uparrow\downarrow}+\mathcal{M}^{21}_{\downarrow\uparrow}\right).
    \end{align}
\begin{widetext}
In the overlap representation of LFWFs, these matrix elements can be written as:
	\begin{align}
	&	\langle P^{\prime}, \lambda^{\prime} | T_g^{11}|P, \lambda \rangle=   \int\frac{{\rm d}^2\bfk {\rm d}x}{16\pi^3 x} \nonumber\\
    &\times  \sum_{\lambda_{s}} \bigg[ \left(2(\kp^{(1)})^{2}+ 2 \kp^{(1)}\qp^{(1)}+i \kp^{(2)}\qp^{(1)}-i \kp^{(1)}\qp^{(2)}\right) \psi_{+1, \lambda_{s}}^{*\lambda^{\prime }}(x,\bfk^{\prime}) \psi_{+1, \lambda_{s}}^{\lambda}(x,\bfk) \nonumber \\
	&+\left(2(\kp^{(1)})^{2}+ 2 \kp^{(1)}\qp^{(1)}-i \kp^{(2)}\qp^{(1)}+i \kp^{(1)}\qp^{(2)}\right)\psi_{-1, \lambda_{s}}^{*\lambda^{\prime }}(x,\bfk^{\prime}) \psi_{-1, \lambda_{s}}^{\lambda}(x,\bfk) 
        \nonumber \\
		& -\left(\frac{(\bfq)^{2}}{2} +i \kp^{(2)}\qp^{(1)}-i \kp^{(1)}\qp^{(2)}\right) \left(\psi_{+1, \lambda_{s}}^{*\lambda^{\prime }}(x,\bfk^{\prime}) \psi_{-1, \lambda_{s}}^{\lambda}(x,\bfk)+\psi_{-1, \lambda_{s}}^{*\lambda^{\prime }}(x,\bfk^{\prime}) \psi_{+1, \lambda_{s}}^{\lambda }(x,\bfk)\right)  
		\bigg],
\end{align}	
\begin{align}
	&\langle P^{\prime}, \lambda^{\prime} | T_g^{21}|P,\lambda\rangle=   \int\frac{{\rm d}^2\bfk {\rm d}x}{16\pi^3 x}  \nonumber\\
    &\times \sum_{\lambda_{s}}\bigg[ \left(2\kp^{(1)}\kp^{(2)}+\kp^{(2)}\qp^{(1)}+\kp^{(1)}\qp^{(2)}\right)\left( \psi_{+1, \lambda_{s}}^{*\lambda^{\prime }}(x,\bfk^{\prime}) \psi_{+1, \lambda_{s}}^{\lambda}(x,\bfk) 
    + \psi_{-1, \lambda_{s}}^{*\lambda^{\prime }}(x,\bfk^{\prime}) \psi_{-1, \lambda_{s}}^{\lambda}(x,\bfk) \right)
    \nonumber \\
	& + i \left(\frac{(\bfq)^2}{2} +i \kp^{(2)}\qp^{(1)}-i \kp^{(1)}\qp^{(2)}\right) \left( \psi_{+1, \lambda_{s}}^{*\lambda^{\prime }}(x,\bfk^{\prime}) \psi_{-1, \lambda_{s}}^{\lambda}(x,\bfk) - \psi_{-1, \lambda_{s}}^{*\lambda^{\prime }}(x,\bfk^{\prime}) \psi_{+1, \lambda_{s}}^{\lambda}(x,\bfk) \right)
	\bigg],
\end{align}	
\begin{align}
	&\langle P^{\prime}, \lambda^{\prime} | T_g^{+1}|P,\lambda \rangle=  P^{+}\int\frac{{\rm d}^2\bfk {\rm d}x}{16\pi^3 }  \nonumber\\
    &\times \sum_{\lambda_{s}}\bigg[ \left(2\kp^{(1)} + \qp^{(1)}-i \qp^{(2)} \right)  \psi_{+1, \lambda_{s}}^{*\lambda^{\prime }}(x,\bfk^{\prime}) \psi_{+1, \lambda_{s}}^{\lambda}(x,\bfk)  + \left(2\kp^{(1)} + \qp^{(1)}+i \qp^{(2)} \right)  \psi_{-1, \lambda_{s}}^{*\lambda^{\prime }}(x,\bfk^{\prime}) \psi_{-1,\lambda_{s}}^{\lambda}(x,\bfk)
    \bigg].
\end{align}
The extraction of $D_{g}(q^{2})$ requires the matrix elements of the $T^{11}$ and $T^{22}$ components of the EMT,  
\begin{align}
    \mathcal{M}^{11}_{\uparrow \downarrow} +
	\mathcal{M}^{22}_{\uparrow \downarrow}+\mathcal{M}^{11}_{\downarrow \uparrow} + \mathcal{M}^{22}_{\downarrow \uparrow} =
	i\Big[B_g(q^2)\frac{q^2}{4M}-D_g(q^2)\frac{3q^2}{4M}+\overline{C}_g(q^2) 2M\Big]\qp^{(2)},\label{c3}
	\end{align}
    which can be expressed in the overlap representation of LFWFs as:
\begin{align}
	&\langle P^{\prime}, \lambda^{\prime} | (T_g^{11}+T_g^{22})|P, \lambda \rangle= 2 \int\frac{{\rm d}^2\bfk {\rm d}x}{16\pi^3 x }   \nonumber \\
	&\times \sum_{\lambda_{s}} \bigg[ \left((\kp^{(1)})^{2}+(\kp^{(2)})^{2}+  \kp^{(1)}\qp^{(1)}+ \kp^{(2)}\qp^{(2)}\right)  \left( \psi_{+1, \lambda_{s}}^{*\lambda^{\prime }}(x,\bfk^{\prime}) \psi_{+1, \lambda_{s}}^{\lambda}(x,\bfk) + \psi_{-1, \lambda_{s}}^{*\lambda^{\prime }}(x,\bfk^{\prime}) \psi_{-1, \lambda_{s}}^{\lambda}(x,\bfk) \right) \nonumber \\
	& + i \left( \kp^{(2)}\qp^{(1)}-\kp^{(1)}\qp^{(2)}\right) \left(\psi_{+1, \lambda_{s}}^{*\lambda^{\prime }}(x,\bfk^{\prime}) \psi_{+1, \lambda_{s}}^{\lambda}(x,\bfk) -\psi_{-1, \lambda_{s}}^{*\lambda^{\prime}}(x,\bfk^{\prime}) \psi_{-1, \lambda_{s}}^{\lambda}(x,\bfk)\right)\bigg]. 
\end{align}
\end{widetext}

%
\bibliography{ref.bib}

\end{document}